# How and why electrostatic charge of combustible nanoparticles can radically change the mechanism and rate of their oxidation in humid atmosphere

Oleg Meshcheryakov[1]

(1) Wing Ltd Company, 33 French boulevard, Odessa, 65000, Ukraine
Email: wing@te.net.ua

**Abstract**

Electrostatically charged aerosol nanoparticles strongly attract surrounding polar gas molecules due to a charge-dipole interaction. In humid air, the substantial electrostatic attraction and acceleration of surrounding water vapour molecules towards charged combustible nanoparticles cause intense electrostatic hydration and preferential oxidation of these nanoparticles by accelerated water vapor molecules rather than non-polar oxygen molecules. In particular, electrostatic acceleration, acquired by surrounding water vapour molecules at a distance of their mean free path from the minimally charged iron metal nanoparticle can increase an oxidative activity of these polar molecules with respect to the nanoparticle by a factor of one million. Intense electrostatic hydration of charged metal nanoparticles converts the nanoparticle's oxide based shells into the hydroxide based electrolyte shells, transforming these nanoparticles into metal/air core-shell nanobatteries, periodically short-circuited by intra-particle field and thermionic electron emission. Partially synchronized breakdowns within trillions of nanoparticles-nanobatteries turn a cloud of charged nanoparticles-nanobatteries – ball lightning – into a powerful radio-frequency aerosol generator. Electrostatic hydration and charge-catalyzed oxidation of charged combustible nanoparticles also contribute to a self-oscillating thermocycling process of evolution and periodic auto-ignition of inflammable gases near to the nanoparticle's surface. The described effects might be of interest for the improvement of certain nanotechnological processes and to better understand ball lightning phenomenon.

# 1. About the possible magnitude and polarity of a net electrostatic charge of ball lightning

Despite numerous attempts, including the most recent ones [1-7], an adequate theoretical and experimental simulation of ball lightning still remains incomplete. At the same time, a simple analysis of the numerous witness descriptions of this phenomenon, carefully collected and classified for example in [8, 9], can provide us with useful information, in particular, concerning the possible magnitude and polarity of a net electrostatic charge of lightning balls.

Some witnesses described a strong attraction of their hair towards lightning balls flying in immediate proximity to them (at distances of about two-three feet). It is interesting that such a strictly directed attraction of human hair to lightning balls with diameters of ~ 10-20 centimetres was repeatedly observed by different witnesses indoors [9]. Our own experience of experimental work, both with highly charged water based artificial clouds and with megavoltage equipment, shows that an attraction of human hair towards such highly charged objects becomes apparent when an average electrostatic intensity reaches ~ 1-2 kV/cm. Therefore, in the above cases one can assume that a potential difference between the visible surfaces of lightning balls and the witnesses' hair could be at least ~ 60-120 kV, and so a net electrostatic charge of such lightning balls could be ~ 1 microcoulomb. The most probable polarity of these lightening balls was negative with respect to the grounded witnesses.

On the other hand, several descriptions from other witnesses of ball lightning phenomenon [9] show that lightning balls can sometimes relatively uniformly and slowly fall from thunderclouds, only appreciably accelerating downwards when approaching the earth's surface. This sudden acceleration, which takes place not far from the earth's surface, and the lightening balls' final elongation to the form of an ellipse before they touch the earth's surface, can indirectly indicate that these balls were charged negatively rather than positively. This is due to the fact that the earth's surface is almost always positively charged with respect to the base of thunderclouds during a thunderstorm.

There are also several detailed descriptions of direct observations of a relatively low-temperature process of ball lightning formation, i.e. ball lightning formation without a previous visible stroke of normal lightning [8, 9]. In particular, such ball lightning formation was repeatedly observed on grounded metal objects, for example, on cast-iron and steel pins of previously destroyed pin-type insulators that were found on pylons of old inoperative electric lines [8, 9]. During a thunderstorm, these grounded rusty pins could probably generate invisible positive streamers. It is possible that simultaneously with the generating of these streamers, the grounded cast-iron and steel pins could electrostatically spray the positively charged iron and carbon based aerosol particles. Such a hypothetical corrosion process accompanied by high-voltage electrospraying of electrohydrated combustible particles from relatively cold iron/carbon based emitters along with a synchronous local generation of a water gas based reducing atmosphere, which form through charge-catalyzed reactions between surrounding water vapour molecules and the rusty iron/carbon surface, could be named 'field-assisted metal dusting corrosion' or 'electrostatic metal dusting' because of its high physical similarity to ordinary metal dusting corrosion [10-16]. At this point, however, it is only important to note that a cloud of the unipolarly charged iron or iron/carbon based combustible nano or micro particles, which are possibly produced in this low-temperature corrosion-electrospraying process, could be a material basis of lightning balls generated from the grounded corona-forming conductors during thunderstorms, and probably such lightning balls could be charged positively rather than negatively.

The positive polarity of the charged lightning balls generated from grounded conductors can also explain their typical horizontal (practically parallel to the earth's surface) flying trajectories at relatively low heights of about 0.5-2 metres above the ground.

In these cases, the identical polarity of the charged lightning balls and the ground might be responsible for a Coulomb repulsion of these balls from grounded objects at short distances.

Having analyzed various witness observations, we can assume that:
(a) the net electrostatic charge of lightning balls can be both positive and negative;
(b) sometimes lightning balls can be exposed to a partial discharging due to either the corona or spark discharges from their surface; the non-contact discharging processes, including those caused by a normal ionic air conduction, can perhaps appreciably reduce the magnitude of an initial electrostatic charge of lightning balls; in particular, such discharging processes can take place when ball lightning approach grounded conducting objects;
(c) sometimes a corona discharge from the surface of highly charged lightning balls can cause a non-contact corona charging of neighbouring low conducting objects with a polarity similar to the polarity of these balls; such a process can cause a subsequent immediate electrostatic repulsion these balls from neighbouring low conducting objects;
(d) a net electrostatic charge of the lightning balls, which attract human hair at distances of about two-three feet, can perhaps reach ~ 1 microcoulomb;
(e) the lightning balls that are frequently described as avoiding contact with grounded conductors and maintaining their approximately constant low flying heights above the ground can be charged with the same polarity as the ground (i.e. positively rather than negatively during thunderstorms), consequently the force of electrostatic repulsion of these balls from the ground and grounded objects at short distances can partially compensate for their weight, contributing to a buoyancy of these balls; a net electrostatic charge of such relatively long-living lightning balls with typical diameters of ~ 10-20 centimetres can be much higher than it was supposed in [17] where an alternative case of unlike charges of ball lightning and the ground was discussed; thus, the magnitude of the net electrostatic charge of lightning balls avoiding contact with grounded conductors can probably also reach ~ 1 microcoulomb, which is in fact limited by a voltage of corona ignition from their surface.

Generally speaking, it was repeatedly assumed, for example in [18, 19], that the high electrostatic charge of lightning balls could play a major role in the existence of ball lightning phenomenon. Equally we share this opinion and in the present paper we will examine the possible role electrostatic charge plays in the life of ball lightning, still assuming that ball lightning is a cloud of combustible aerosol particles that are exposed to a slow, predominantly electrochemical oxidation [20]. Such a process of the electrochemical oxidation of nano or submicron aerosol particles converts these combustible particles into aerosol batteries - further, for short 'nanobatteries' - that are periodically (and perhaps with very high frequencies) short-circuited by intra-particle breakdowns.

According to [20]:

    (a) The aerosol particles-batteries can exist either in the form of nano or submicron aggregates, or in the form of nano or submicron core-shell capsules, or in a more realistic combination of these two simplest types, i.e. in the form of highly aggregated aerosol structures consisting of mixed nano and submicron core-shell particles;

    (b) These aerosol particles-batteries can contain at least one reductant component, e.g. a metal or carbon based reductant component, and at least one electrolyte component;

    (c) The aerosol particles can use either an internal compact oxidant or external oxidant from ambient air, i.e. oxygen gas and/or water vapour;

    (d) During the process of the electrochemical oxidation the aerosol particles automatically turn into either aggregated or core-shell structured aerosol nanobatteries periodically short-circuited by the intra-particle breakdowns due to both field and thermoionic electron emission taking place within and on the surface of particles-batteries;

    (e) The short-circuited aerosol particles-nanobatteries are free magnetic dipoles, and so they can be exposed to an intense mutual magnetic dipole-dipole attraction, forming ball-shaped self-assembling clouds with high magnetic polarizability;

    (f) The non-short-circuited aerosol particles-nanobatteries are free electric dipoles with substantial electric dipole moments, and so they can be exposed to an intense mutual electric dipole-dipole attraction, forming ball-shaped self-assembling clouds with high electric polarizability (Fig. 1);

    (g) The repeating processes of the short circuits within the separate aerosol particles-nanobatteries can be partially or totally synchronized within a ball lightning cloud; such repeating synchronized collective short circuits of trillions of nano or submicron batteries can generate powerful electromagnetic radiation, for example, in microwave range, which, in particular, could explain the repeatedly observed cases of intense red heat of incandescent bulbs' filaments which are switched off from power sources; such a temporary intense red heat of the filaments sometimes was distantly induced during the slow flying of ball lightning at distances of about one-two feet from the switched off bulbs [9].

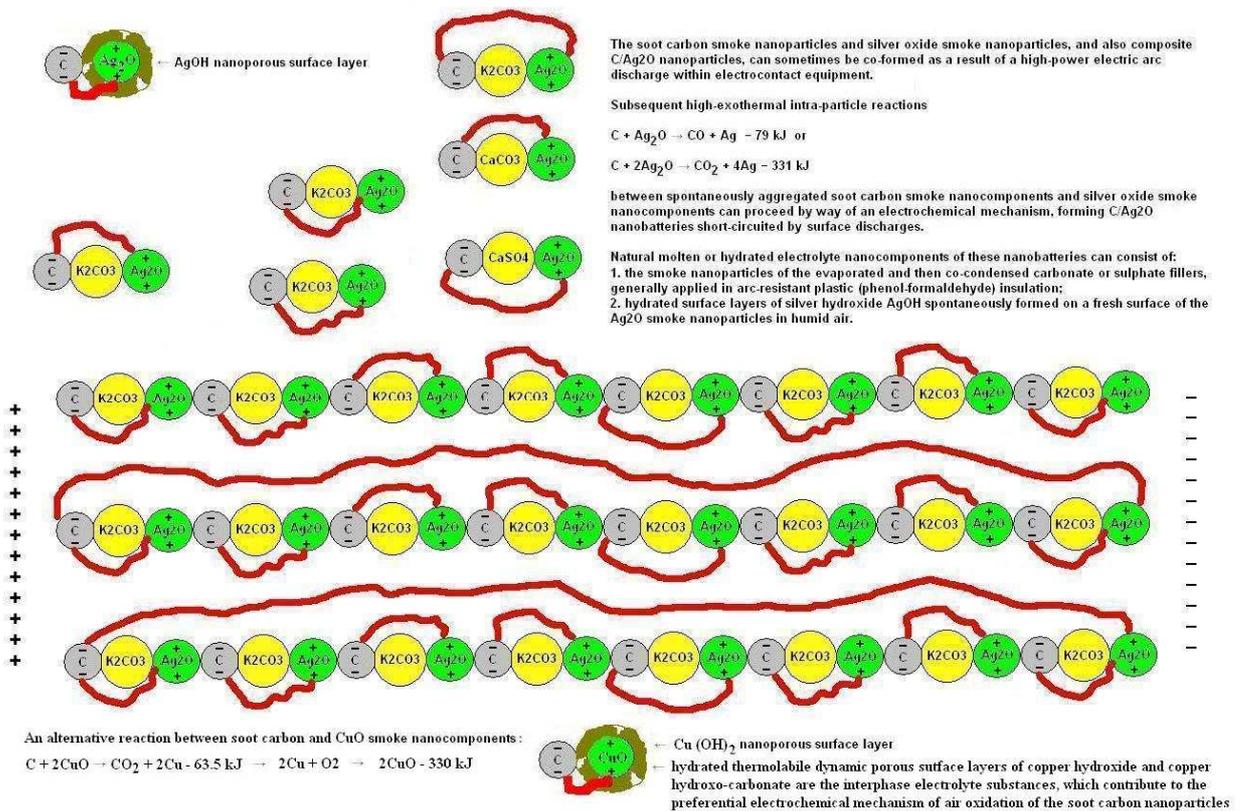

Figure 1. External electrostatic fields and the high initial concentration of aggregated aerosol nanoparticles-nanobatteries facilitate electrostatic aggregation of individual nanobatteries and automatic conversion of these composite aerosol aggregates to microscopic nanoparticle based high-voltage generators. Ball lightning can contain some hundreds of billions of individual nanobatteries per cubic centimetre. Every such individual nanobattery is an electric dipole and every short-circuited individual nanobattery is also a magnetic dipole. If the initial concentration of the aerosol nanobatteries is high enough, these nanobatteries can spontaneously form metastable micrometre-sized aerosol chains-aggregates due to the mutual electrostatic dipole-dipole attraction. However, an external electrostatic field, for example a thunderstorm electrostatic field, can modify this self-assembling process. The spontaneous or externally induced electrostatic aggregation of the nanobatteries enables the periodic formation of relatively long electric circuits consisting of many nanobatteries temporarily connected in series. The micrometre-sized chain aggregates, which contain only ten thousand individual particles-nanobatteries, can locally generate a ten kilovolt voltage with corresponding macroscopic spark discharges. In particular, the aggregated composite aerosol nanobatteries can be spontaneously formed from soot aerosol nanoparticles - nanoanodes and metal oxide oxidant nanoparticles - nanocathodes. Aerosol nanobatteries-nanoaggregates consisting of the electrostatically unipolar charged soot carbon reductant nanoparticles and CuO solid oxidant nanoparticles (or also of the soot carbon reductant nanoparticles and $Ag_2O$ oxidant nanoparticles) could be one of the most probable components of lightning balls frequently generated from electrical equipment during thunderstorms.

When discussing the simplest possible processes of spontaneous formation of the short-circuited aerosol nanobatteries from combustible aerosol nanoparticles in a storm atmosphere, it was assumed in [20] that a high concentration of water vapour in the air can significantly modify the mechanism of oxidation of many metal aerosol nanoparticles, converting a normal process of direct oxidation of these nanoparticles by neutral oxidizing species into the predominantly electrochemical, i.e. ion-mediated oxidation process in this highly humid atmosphere. During such an electrochemical oxidation, water vapour from humid air contributes to the formation of the hydrated, hydroxide or oxy-hydroxide based, thermolabile porous layers on the surface of metal nanoparticles rather than contributing to the formation of more thermostable layers of mixed anhydrous metal oxides [19], normally growing on the surface of nanoparticles either at very high, dehydrating temperatures or only in dry oxygen [21-26].

In humid air, a quick diffusion transfer of ions can take place through the hydroxide based electrolyte layers growing on the surface of the nanoparticles due to their oxidation by water vapour. The quick processes of the diffusion of ionized oxidizing species, such as ions $OH^-$, or $O_2^-$, or $CO_3^{2-}$, that easily migrate through the dynamic surface electrolyte layers of the mixed hydroxides towards the metal core of the nanoparticle, can considerably prevail above the much slower alternative processes of the inward diffusion of neutral oxidizing species, such as molecular $O_2$, or $H_2O$, or $CO_2$, or neutral radicals like $OH\cdot$.

Similarly, the quick processes of the diffusion of ions of metal core, such as $Fe^{2+}$, or $Fe^{3+}$, or $Fe(OH)^+$, $Fe(OH)^{2+}$, to the nanoparticle surface also can significantly prevail above much slower alternative processes of diffusion of neutral atoms from the metal core to the surface of the nanoparticle.

Thus, at relatively low temperatures, while electrolyte layers growing on the surface of metal nanoparticles remain thermostable, a process of the water vapour induced electrochemical oxidation of combustible nanoparticles can become a most preferable process of their oxidation in humid air because of the possibility of the quick intra-particle transport of ions, and particularly because of the possibility of the quick transport of ionized oxidizing species through the hydrated surface electrolyte layers towards oxidable cores of the combustible nanoparticles.

It seems that in addition to high air humidity of another extremely important complementary condition is necessary in order that the process of electrochemical oxidation of combustible aerosol nanoparticles can substantially prevail over the alternative process of their normal oxidation by neutral oxidizing species from ambient air.

We suppose that such a complementary condition, radically changing the process of oxidation of combustible nanoparticles in humid air, can be a high electrostatic charge of these nanoparticles.

Further it will be shown how and why a presence of electrostatic charges on combustible aerosol nanoparticles can make an important contribution to their preferential oxidation by water vapour molecules but not by much more numerous oxygen gas molecules in humid air, i.e. how and why electrostatic charges distributed on the surface of combustible aerosol nanoparticles can become powerful selective catalysts of water vapour induced oxidation of these nanoparticles.

## 2. Electrostatic hydration of atmospheric ions and charged aerosol nanoparticles

As is well known, in normal, humid air, intense charge-dipole interaction between atmospheric ions and highly polar molecules of water vapour causes immediate hydration of the ions. A hydrated ion includes a central ion and a water shell normally consisting of several $H_2O$ molecules. The hydrated ions are extremely stable because of the huge electrostatic energy which keeps polar water molecules in immediate proximity to the central ion (the typical energies of complete dehydration of atmospheric ions can be ~ several electron-volts). Therefore the hydrated ions are a standard form of gaseous ions in the lower troposphere [8, 18, 27]. The charge-dipole interaction between gaseous ions and surrounding polar gas molecules is a powerful and long-range attraction, and consequently processes of intense hydration of gas ions caused by the ion-dipole attraction can easily find useful scientific applications, for example for an effective operation of the Wilson chamber where gaseous ions generated by ionizing radiation act as condensation nuclei in order to visualize tracks of ionizing particles. In this well known case, hydrated or alcohol-solvated gaseous ions quickly turn into water or alcohol based electrostatically charged nanoparticles. These charged nanoparticles quickly grow into micrometre-sized droplets due to their further intense electrostatic hydration/ solvation under supersaturation conditions. Electrostatically charged aerosol nanoparticles, especially 'small' nanoparticles with characteristic sizes of about several nanometres, almost do not differ from gaseous ions in their ability to attract surrounding polar gas molecules, including water vapour molecules, from ambient atmosphere due to the powerful long-range charge-dipole interaction. In humid air, a surplus electrostatic charge of such nanoparticles is almost always localized to trapping sites on the nanoparticles surface (in the form of either one or several adsorbed hydrated ions), i.e. very close to surrounding polar gas molecules.

If the charged nanoparticles are conducting, surplus electrostatic charges can migrate around their surface in a random manner. In addition, both the conducting and non-conducting charged aerosol nanoparticles can freely and irregularly revolve on their axes due to continuous stochastic Brownian collisions with surrounding gas molecules. In this case, a time-average density of surplus electrostatic charge on the nanoparticle surface can be practically equivalent to the time-average charge density on the surface of the same charged nanoparticle whose surplus electrostatic charge is localized in its centre, i.e. the surplus electrostatic charge is quasi-distributed on the surface of the nanoparticle.

One way or another, it is very likely that the surplus charges distributed on the surface of the hydrophobic or hydrophilic nanoparticles will make a major contribution to their local or total surface hydration in humid air due to the powerful charge-dipole attraction of surrounding polar molecules of water vapour, which is very similar to the hydration of atmospheric ions mentioned above.

In fact, in this case the charge-dipole attraction of water vapour molecules towards the charged surface of the nanoparticle plays the role of a powerful electrostatic water pump.

If the charged nanoparticle is cold enough, the electroadsorbed water molecules can strongly hold on to the cold charged surface of the nanoparticle.

If the charged nanoparticle is heated to a high enough temperature, the water vapour molecules previously electroadsorbed on the charged surface of this nanoparticle can be thermally desorbed when heating.

If the electroadsorbed water molecules are not consumed on the surface of the cold charged nanoparticle, for example due to some possible surface reactions, the surface of the charged nanoparticle will remain highly hydrated till the nanoparticle is cold enough.

If the charged nanoparticle consists of a substance that can be oxidized by water vapour at a given temperature, the water vapour induced oxidative reactions will inevitably take place on the surface of such a charged nanoparticle.

In many cases, during water vapor induced oxidation of combustible nanoparticles, for example metal nanoparticles, a complete cascade of the primary and secondary oxidative reactions accompanying the water vapour induced nanoparticle oxidation can be highly exothermal, and so, because of the series of such highly exothermal oxidative reactions, a temperature of the charged 'electrohydrated' nanoparticle will grow to some limiting value, while the electroadsorbed water vapour molecules will be subject to thermal desorption from the heated nanoparticle surface, despite the continuous functioning of the 'electrostatic water pump'. The thermal desorption of the water molecules will reduce the rate of the water vapour induced oxidative surface reactions, consequently the charged nanoparticle will quickly cool, and so the process of the electrostatic, charge-dipole adsorption of the water vapour molecules will recommence. Again the rate of the exothermal oxidative reactions will grow, and it will again cause the growth of temperature of the charged nanoparticle and so on, and so forth. This cyclical oxidative process will repeat until the combustible charged nanoparticle is completely oxidized. Probably a frequency of such a self-oscillating oxidative process might be very high, as possible rates of heating/cooling of the nanoparticles are extremely high.

Clearly, a balance between the competing processes of the electrostatic oxidative adsorption of water vapour molecules on the surface of the charged nanoparticle and thermal desorption of water molecules from this surface can be achieved either in a mode of a synchronous running of both these processes, or in the self-oscillating mode of a successive alternation of these processes, or in a combination of these two modes. It seems, however, that the self-oscillating mode of the successive alternation of the processes of the electrostatic oxidative adsorption of the water vapour molecules on the surface of the charged nanoparticle and their thermal desorption from this surface would be one of the most probable modes. The additional reasons for the high probability of such a self-oscillating mode of water vapour induced oxidation of combustible charged nanoparticles will be discussed in the next section.

## 3. The mechanisms and products of water vapour induced oxidation of combustible nanoparticles radically differ from the mechanisms and products of their oxidation by oxygen gas

At least two main gas oxidants with substantial partial pressures are available in the air to oxidize combustible aerosol nanoparticles, irrespective of whether these nanoparticles are electrostatically charged, and consequently they can actively electroadsorb polar molecules of water vapour from ambient air, or these nanoparticles are electrostatically neutral, and consequently they will be much more indifferent to surrounding water vapour.

These two main atmospheric oxidants are:

    (a) oxygen gas, $O_2$, with a mole fraction of oxygen molecules in the air at Sea level, $n_O$, ~ 0.21 (i.e. ~ 0.21 mol of oxygen gas per one mol of air);

    (b) water vapour, $H_2O$, with a mole fraction of water molecules, $n_W$, ranging in the air at Sea level between ~ 0.002 (at very low air temperatures, in particular, in Antarctica) to ~ 0.01 - 0.02 (at normal summer temperatures in temperate latitudes), to ~ 0.03 air humidity maximum (in the tropics or during summer thunderstorms in temperate latitudes).

In this paper we will use the term 'humid air' for normal air atmosphere where a mole fraction of molecules of water vapour, $n_W$, ranges from 0.01 to 0.03 mol of water vapour per one mol of air.

Thus, the term 'humid air' will be in fact equivalent to the term 'normal air', as the humidity of such 'humid air' is typical not only for most thunderstorms, but also practically for any summer weather.

In this paper, we also conditionally use the term 'combustible nanoparticles', which requires a more precise definition. This term is used by us to denote aerosol nanoparticles or substrate-integrated/ substrate-precipitated nanostructures consisting of condensed materials, which are able to be oxidized by surrounding molecules of both oxygen gas and water vapour, irrespective of the specific mechanism, rate and optimal temperature of such oxygen gas or water vapour induced oxidation.

And so the term 'combustible nanoparticles' can be, in particular, applied to aerosol nanoparticles or substrate-integrated/ substrate-precipitated nanostructures consisting of the overwhelming majority of metals, metalloids, intermetallides, sulfides, hydrides, carbides, phosphides, nitrides, silicides, borides, selenides, tellurides, lower oxides, many organic compounds, particularly unsaturated organic compounds, many polymer and biopolymer structures. A lot of carbon based nanoparticles, either aerosol or precipitated ones, for example such as soot nanoparticles, or fullerenes, or carbon nanotubes can also be considered as 'combustible nanoparticles' because they can be theoretically oxidized by both surrounding oxygen gas molecules and/or water vapour molecules. Thus, in this paper, a wide range of natural and man-made nanoparticles will be conditionally considered as 'combustible nanoparticles' in the above-mentioned aspect.

The mechanisms of oxidation of combustible aerosol nanoparticles by each of the two competing atmospheric oxidants are radically differing.

As mentioned above, reactions of the dry oxygen oxidation of many combustible (e.g. metal or metalloid) aerosol nanoparticles can give reaction products in the form of solid or molten layers of mixed oxides, growing on the surface of the nanoparticles during the process of their oxidation.

Alternatively, reactions of oxidation of the combustible aerosol nanoparticles by molecules of pure water vapour usually give simultaneously two types of different reaction products:

a) solid reaction products in the form of the more or less hydrated, more or less thermostable, more or less porous metal hydroxide shells on the surface of the nanoparticles, and in addition

b) combustible gases, in particular, hydrogen gas, when oxidizing the nanoparticles of the great number of reactive metals or metalloids, for example, when oxidizing the nanoparticles of such different substances as aluminium, iron, tungsten, molybdenum, zirconium, calcium, cadmium, silicon etc.

As an example, let us compare the reaction products synthesized when oxidizing the silicon based aerosol nanoparticles either in dry oxygen, or in pure water vapour, or in humid air.

Aerosol nanoparticles that consist of pure silicon can be oxidized in dry oxygen to generate nanolayers of the mixed silicon oxides growing on their surface:

$$\text{Si}_{\text{(nanoparticle core)}} + \text{O}_2 = \text{SiO}_{2\,\text{(surface nanolayer)}} \text{ (dielectric)} \tag{1}$$

$$2\text{Si}_{\text{(nanoparticle core)}} + \text{O}_2 = 2\text{SiO}_{\text{(surface nanolayer)}} \text{ (volatile combustible dielectric)} \tag{2}$$

At high temperatures (~ 700-1300°C), either in pure water vapour or in humid air the silicon aerosol nanoparticles can be oxidized to generate growing surface layers of the mixed dielectric silicon oxides plus the evolved hydrogen gas:

$$\text{Si}_{\text{(nanoparticle core)}} + 2\text{H}_2\text{O}_{\text{(vapour)}} = \text{SiO}_{2\,\text{(surface nanolayer)}} \text{ (dielectric)} + 2\text{H}_2\uparrow \tag{3}$$

$$\text{SiO}_{\text{(surface nanolayer)}} + \text{H}_2\text{O}_{\text{(vapour)}} = \text{SiO}_{2\,\text{(surface nanolayer)}} \text{ (dielectric)} + \text{H}_2\uparrow \tag{4}$$

At relatively low temperatures, in humid air products of the water vapour induced oxidation of silicon nanoparticles can include the evolved hydrogen gas and mixed layers of silicic acids, growing on the nanoparticles surface instead of mixed anhydrous silicon oxides. These dynamic porous surface nanolayers can consist of either metasilicic acid $\text{H}_2\text{SiO}_3$, or orthosilicic acid $\text{H}_4\text{SiO}_4$, or disilicic acid $\text{H}_2\text{Si}_2\text{O}_5$, or pyrosilicic acid $\text{H}_6\text{Si}_2\text{O}_7$, or of the mixed silicic acids. Such hydrated surface nanolayers are water-soluble electrolytes with a relatively low thermal stability and low ionic conduction, which, nevertheless, can probably make some contribution to the process of low temperature electrochemical oxidation of silicon nanoparticles [20]:

$$\text{Si}_{\text{(nanoparticle core)}} + \text{H}_2\text{O}_{\text{(vapour)}} + \text{O}_2 = \text{H}_2\text{SiO}_3 \text{ (water-soluble volatile electrolyte)} \tag{5}$$

$$\text{SiO}_{2\,\text{(surface nanolayer)}} + \text{H}_2\text{O}_{\text{(vapour)}} = \text{H}_2\text{SiO}_3 \text{ (water-soluble volatile electrolyte)} \tag{6}$$

Silicon-containing aerosol nanoaggregates, which in addition to nanoparticles of pure silicon contain some mineral nanocomponents, such as co-aggregated nanoparticles of either NaOH, or KOH, or Ca(OH)$_2$, or Ba(OH)$_2$, can be co-condensed for example from the plasma-evaporated silicate minerals in a carbon monoxide reducing local atmosphere, which is practically identical with the process described in [19].

In humid air products of the water vapour induced oxidation of such silicate-containing aerosol nanoaggregates, can include evolved hydrogen gas and nanolayers of thermostable water-soluble and/or molten silicate electrolytes:

$$\text{Si}_{\text{(nanoparticle)}} + 2\text{NaOH}_{\text{(mineral nanoimpurity)}} + \text{H}_2\text{O}_{\text{(vapour)}} = \text{Na}_2\text{SiO}_3 \text{ (electrolyte)} + 2\text{H}_2\uparrow \tag{7}$$

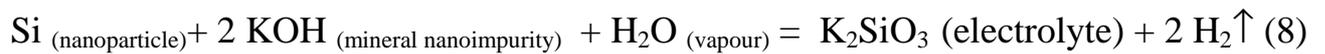

Si (nanoparticle) + 2 KOH (mineral nanoimpurity) + $H_2O$ (vapour) = $K_2SiO_3$ (electrolyte) + 2 $H_2\uparrow$ (8)

Thus, the described above example of the possible alternative pathways of the dry oxygen and humid air induced oxidation of the silicon or silicon based nanoparticles illustrates a simple fact: in humid air, reactions of oxidation of combustible aerosol nanoparticles by water vapour frequently can be accompanied by an evolving of combustible gases. Such an evolving of combustible gases can take place through the process of water vapour induced oxidation of very different nanoparticles; and this evolving of the combustible gases in turn can be accompanied by their spontaneous ignition in humid air.

Within a small cloud of predominantly unipolarly charged red-hot nanoparticles-nanobatteries, i.e. within ball lightning, there are several potential reasons for the continuous or repeating processes of spontaneous ignition of evolved combustible gases. For example, such a spontaneous ignition of combustible gases can occur due to the repeating collective discharge processes of short circuits taking place within and on the nanoporous surface of the aggregated aerosol nanobatteries. On the other hand, such repeating processes of auto-ignition of the evolved combustible gases can occur due to contact of the evolved gases with a fraction of micrometre-sized permanently red-hot aerosol particles - 'permanent aerosol igniters' - that in turn were previously heated by the preceding intra-cloud exothermal oxidative reactions, including for example oxygen gas induced oxidative reactions; these substantially slowed oxygen induced oxidative reactions can still proceed in humid air on a relatively low-hydrated surface of a fraction of minimally charged or uncharged particles co-aggregated with highly charged particles in order to form composite charged aerosol aggregates, perhaps constituting ball lightning.

It would be reasonable to assume that if ball lightning really is a cloud of charged combustible particles-nanobatteries subjected to slow, predominantly electrochemical oxidation, this cloud most probably contains an extremely polydisperse ensemble of highly aggregated and variously charged particles. One hypothetical fraction of aerosol particles constituting ball lightning could include nano or submicron particles positively charged with possible charge limits of ~ 1 up to ~ 10 surplus elementary charges per one nano/submicron particle. Another fraction of aerosol particles, which initially form ball lightning cloud, could include nano or submicron particles negatively charged with the same charge limits. Gradual electrostatic aggregation of the first and second aerosol fractions can result in gradual mutual recombination of opposite charges within ball lightning. As mentioned above, the positive and negative charges are substantially unbalanced in ball lightning, totally generating a net electrostatic charge of ball lightning that can reach ~ 0.1-1 microcoulomb. A third hypothetical fraction of combustible aerosol particles constituting ball lightning could contain minimally charged or neutral nano, submicron or micrometre-sized aerosol particles, which at the same time can be polarized and aggregated together in the form of chains [19] by both a charge-dipole and dipole-dipole electrostatic interaction. A fourth hypothetical fraction of aerosol particles constituting ball lightning could include relatively large, micrometre-sized highly charged aerosol particles, which play the role of local electrostatic aerosol collectors precipitating numerous surrounding chain aerosol aggregates consisting of bipolarly minimally charged or polarized nano and submicron particles. A large surface of such a micrometre-sized aerosol particle-collector can be charged with tens or even hundreds of surplus elementary charges (i.e. adsorbed hydrated ions). These charges, relatively evenly distributed on the surface of the micrometre-sized particles, are local centres of intense electrostatic attraction of the surrounding highly polarized chain aerosol aggregates. Electrostatic precipitation of the surrounding minimally oppositely charged aggregated chains radially directed to the surface of the micrometre-sized charged particles-collectors, can contribute to a formation of the sea-urchin, dandelion-, or hedgehog-like structures, such as those described in [6].

Surplus charges, distributed on the surface of the micrometre-sized charged particles-collectors, are centres of electrostatic attraction not only for surrounding nano and submicron chain aerosol aggregates but also for surrounding polar molecules of water vapour. Therefore the highly charged surface of such micrometre-sized particles can play the role of a catalytic surface where the combustible nano and/or submicron particles constituting chain aggregates will consecutively react with electroadsorbed (and electrostatically accelerated) molecules of water vapour (Fig. 2). In this case, charges permanently fixed on the surface of the micrometre-sized particle-collector, for example in the form of adsorbed ions such as $H_3O^+ (H_2O)_n$, $OH^- (H_2O)_n$ or $O_2^- (H_2O)_n$, can play the role of catalysts of local water vapour induced oxidation of combustible nano and submicron sized minimally charged or neutral, and only polarized, aerosol particles.

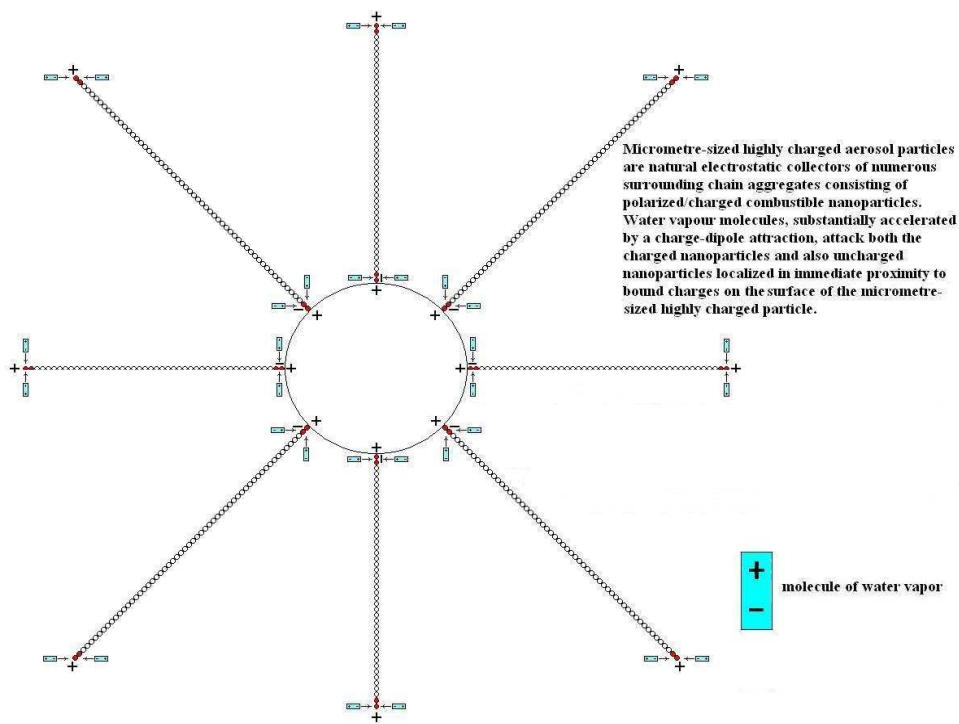

Figure 2. Radius of the nanoparticle surface zone, most intensively attacked by surrounding electrostatically accelerated water vapor molecules, can reach ~ 1-2 nm in immediate proximity to the site where the surplus charge (i.e. adsorbed hydrated ion) is located at this moment.

So, during the electrostatic precipitation of the chains of the aggregated nano and submicron aerosol particles on the charged surface of the large particles-collectors, the combustible particles from these chains can be exposed to intense water vapour induced oxidation, locally catalyzed by the surplus electrostatic charges fixed on this surface. Such a scenario of the water vapour induced oxidation of the minimally bipolar charged or even neutral but polarized combustible nanoparticles really could require the presence of the fourth hypothetical fraction consisting of the micrometre-sized highly charged aerosol particles – natural catalytically active electrostatic aerosol collectors of both water vapour and surrounding combustible nanoparticles.

Alternatively, surplus electrostatic charges, capable of catalyzing the water vapour induced oxidation of combustible nanoparticles, could move along aerosol chain aggregates, step by step contributing to the consecutive oxidation of nanoparticles within chains. Such a process of consecutive charge-catalyzed water vapour induced oxidation of aggregated combustible nanoparticles by a stepped shift of surplus charge along the chain of the aggregated nanoparticles can resemble the process of Bickford's fuse combustion (Fig. 3, 4).

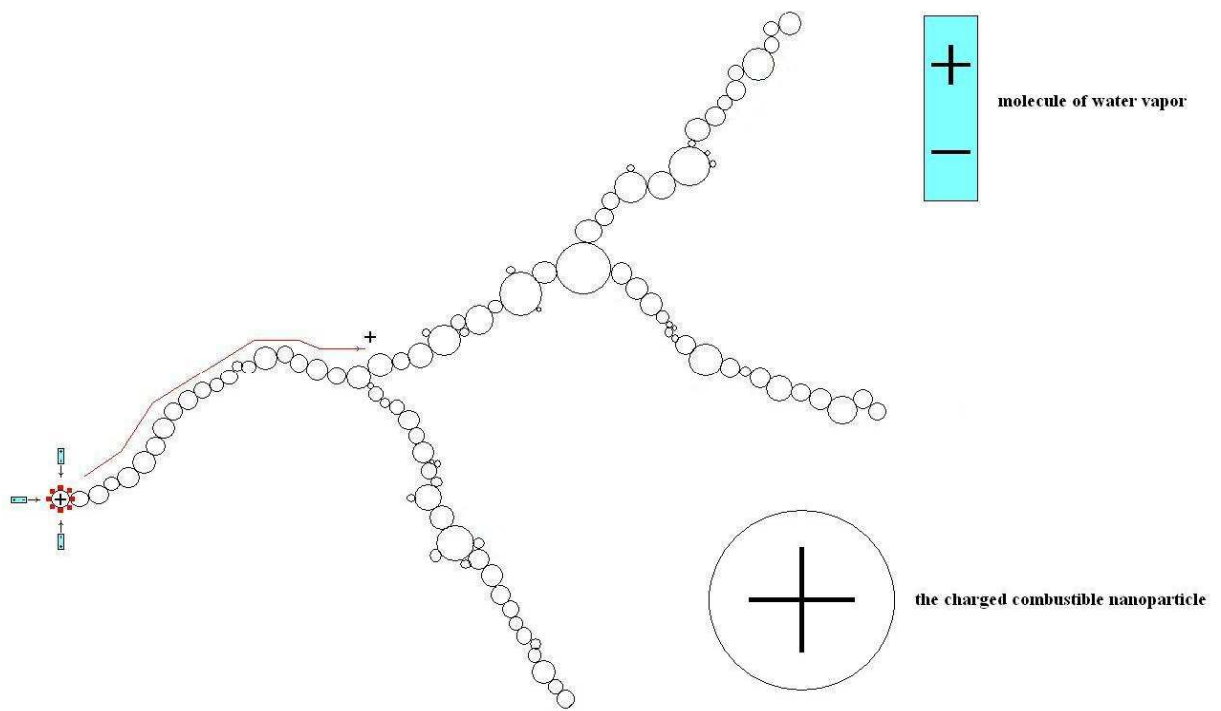

Figure 3. Stepped movement of an electrostatic charge (for example, a hydrated ion) along a chain of combustible aerosol nanoaggregates successively catalyzes water vapour induced oxidation of these nanoparticles, step-by-step burning the nanoaggregates.

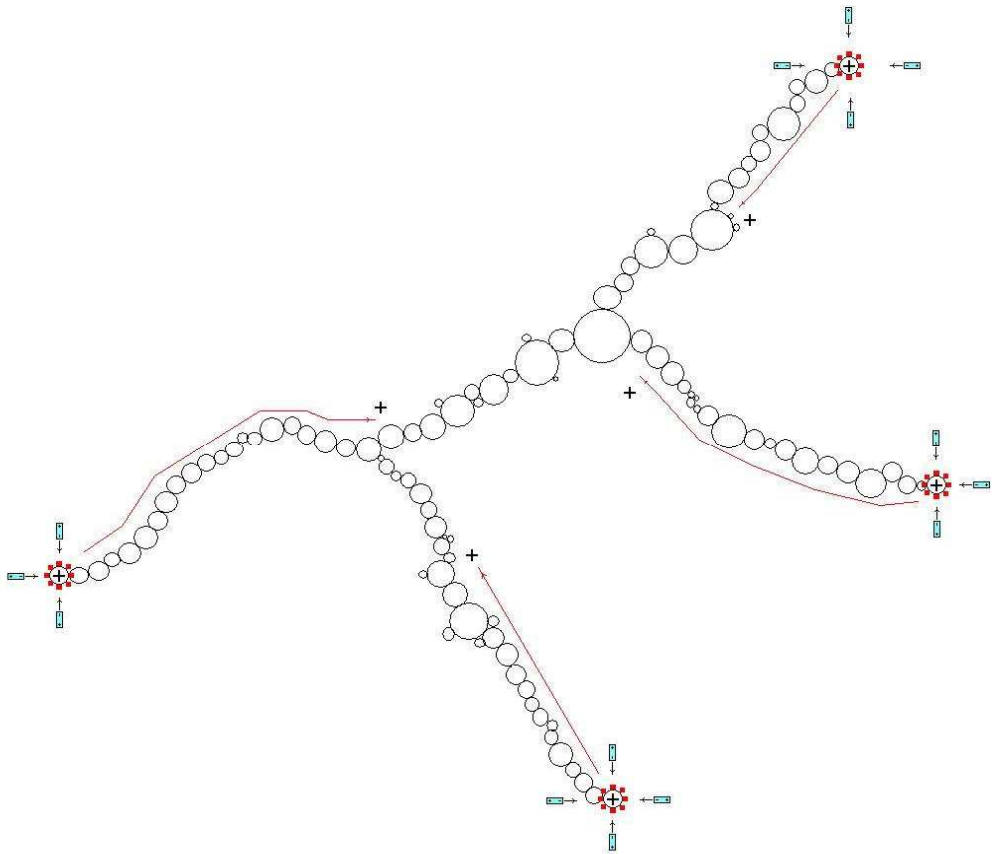

Figure 4. Branched chain aggregates of combustible soot or metal nanoparticles are a frequent aerosol product of high-temperature condensation of carbon and metal vapour evaporated by an electric arc. These chain aggregates can contain a lot of mutually linked small nanoparticles, and every such micrometre-sized aerosol aggregate can be charged (with either an elementary charge or with several elementary charges). An elementary charge located on a peripheral nanoparticle in the chain is a powerful point catalyst of its oxidation by surrounding polar molecules of water vapor. When the first peripheral nanoparticle is entirely oxidized, the elementary charge jumps onto the next nanoparticle in the chain catalyzing its water vapour induced oxidation, then the process repeats again etc. During such a charge-catalyzed successive oxidation of nanoparticles in aerosol aggregate chains, the elementary charge moves along the chain of the oxidizable nanoparticles linked with a wave of water vapor induced oxidation of the nanoparticles in a similar way to how a flame front moves along a Bickford's fuse. Thus, even a single elementary charge can enable the successive catalytic oxidation of a lot of the small combustible nanoparticles, when this charge moves along the aerosol nanoaggregate chains. The stepped movement of several electrostatic charges (for example, unipolar charged hydrated ions) along the branched chains of combustible aerosol nanoaggregates successively catalyzes water vapour induced oxidation of these nanoparticles, ultimately, completely burning off the nanoaggregates.

Clearly, if characteristic sizes of a combustible particle are of ~ 10-1000nm or greater, unipolar surplus charges (i.e. adsorbed hydrated ions) distributed on the surface of such a particle can form only local mobile surface spots of extremely high electrostatic intensity. The charge-dipole interaction will cause intense attraction and even local acceleration of the surrounding water vapour molecules to these highly charged surface spots, consequently such intense electrostatic hydration of the charged spots on the surface of a combustible particle will result in active water vapour induced oxidation of these small charged sites of the surface of a combustible particle. Being thermally activated during the oxidative process, the surplus electrostatic charges

(adsorbed hydrated ions) can jump on the surface of the heated particle in a random manner. And so, this consecutive electrostatic water vapour induced 'charge spot corrosion' of the nano-, submicron- or micrometre-sized particles step-by-step can oxidize even such large charged particles - particularly when their surface is highly hydrophilic and so locally electroadsorbed water vapour molecules can be next distributed on all the surface of the particle by wetting (Fig. 5).

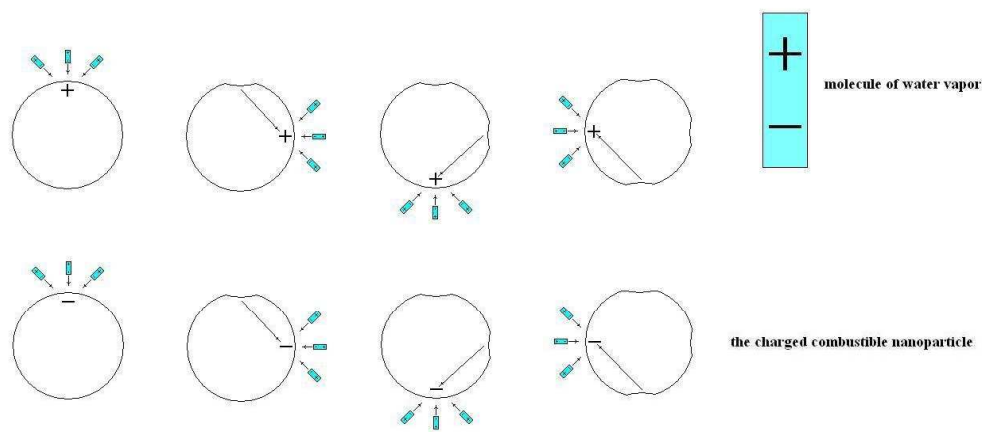

Figure 5. Successive 'jumping' of the surplus charge – a local catalyst of water vapour induced oxidation of combustible nanoparticle - between the charge trapping sites on the surface of the relatively large (i.e., ~ 5 - 100 nm in diameter) combustible nanoparticle causes numerous successive events of local corrosion of this surface with a final water vapour induced oxidation of the whole nanoparticle.

Thus, a water vapour induced oxidation of combustible nanoparticles, and particularly the charge-catalyzed water vapour induced oxidation of combustible nanoparticles, is almost always accompanied by an evolution of combustible gases.

Only when a concentration of the evolved combustible gases reaches a lower flammability limit within the nanoparticles cloud, the gases can locally ignite, and a radially spreading deflagration wave can then ignite the whole cloud. Such high temperature deflagration waves can periodically propagate within the cloud of combustible aerosol nanoparticles and contribute to the periodic thermal dehydration of the electrohydrated nanoparticles. This periodic thermal dehydration of the combustible nanoparticles can temporarily interrupt their water vapour induced oxidation. However if aggregated combustible aerosol nanoparticles constituting this small cloud contain a fraction of charged nanoparticles, the powerful electrostatically induced adsorption of polar molecules of water vapour from ambient humid air will continuously re-hydrate these charged nanoparticles, contributing to reactivation of their water vapour induced oxidation, as well as probably to an intra-cloud cyclization and synchronization of the successive processes of electrostatic oxidative hydration and thermal dehydration of the charged combustible nanoparticles.

Each repeating cycle of such intra-cloud collective processes can consist of successive stages:

(a) electrostatic oxidative hydration of the charged combustible nanoparticles accompanied by evolving combustible gases;

(b) auto-igniting the evolved gases, for example by their contact with a fraction of the permanently red-hot 'igniter' micrometre-sized aerosol particles;

(c) thermal dehydration of the majority of the combustible nanoparticles.

Within a ball lightning cloud, these three-stage cycles can recur repeatedly in the form of a self-oscillating process. An intra-cloud synchronization of the repeating stages of water vapour induced oxidation of charged combustible nanoparticles is probably only one of the possible scenarios, but a tendency to such an intra-cloud thermodynamic stage synchronization can arise from at least six interconnected circumstances: (1) relatively slow evolving combustible gases within the ball lightning cloud; (2) relatively fast intra-cloud propagation of the deflagration waves, which consequently can contribute to practically synchronous thermal dehydration of almost all the nanoparticles within the cloud; (3) extremely low thermal inertia of the majority of nanoparticles within the cloud; (4) relatively high thermal inertia of the small fraction of micrometer sized permanently red-hot aerosol particles-igniters within the cloud; (5) existence of the lower flammability limits of evolved combustible gases; (6) existence of the relatively prolonged ignition delay time, in particular, for ignition in hydrogen/air mixtures by high-frequency streamer discharges [28].

If the gas evolved by charged combustible nanoparticles-nanobatteries during the process of their water vapour induced oxidation is hydrogen, this combustible gas can be periodically autoignited either by collective discharge short circuits within and on the surface of the aerosol particles-nanobatteries or by contact of the hydrogen with the fraction of the large, micrometre-sized aerosol particles-igniters permanently heated to a temperature higher than ~ 585°C, which is the autoignition temperature of air-hydrogen mixtures, or even at a lower temperature - as a result of potential catalytic effects of the highly developed oxidized surface of the heated nanoparticles (for example such as iron based nanoparticles, which can substantially catalyze the air-hydrogen oxidation reactions).

If the cloud of metal aerosol nanoparticles is exposed to a preferential oxidation by water vapour, the evolving hydrogen can be either autoignited or non-autoignited depending on the concrete local temperature and concentration conditions.

The evolving hydrogen gas can react with ambient atmospheric oxygen both directly on the surface of the metal aerosol nanoparticles and partially in the surrounding gas phase. When the evolving hydrogen is autoignited by oxygen from ambient humid air, we do not see a flame directly, because the air-hydrogen flame is visible only in ultra-violet and not in visual range.

However, the ball lightning cloud of electrostatically charged red-hot nanoparticles can play the role of a relatively low-quality natural aerosol visualizer of the air-hydrogen flame generated by this cloud during the process of the preferential water vapour induced oxidation of the charged metal nanoparticles. If the combustible aerosol nanoparticles are strongly interconnected within the cloud by a long-range mutual attraction, for example by the dipole-dipole magnetic attraction, they can form a relatively heavy stable ball-shaped cloud, and so convection currents produced by the air-hydrogen flame can not appreciably influence both the motion and shape of this heavy stable aerosol cloud [20].

Thus, in this case we can only see a luminous ball-shaped cloud of the distantly interconnected red-hot aerosol nanoparticles, but not an air-hydrogen flame cone.

In addition to hydrogen, the different combustible gases can also be generated through the process of the water vapour induced oxidation of different combustible aerosol nanoparticles, including non-metallic ones.

For example, a combustible mixture of carbon monoxide and hydrogen, so-called water gas, can be evolved during the process of preferential water vapour induced oxidation of carbon based nanoparticles, such as soot nanoparticles, or fullerenes, or carbon nanotubes, or metal carbide nanoparticles. A completed water vapour induced oxidation of these nanoparticles will generate gas reaction products in the form of carbon dioxide and combustible hydrogen. Alternatively, the completed dry oxygen induced oxidation of these nanoparticles would generate only noncombustible carbon dioxide.

Similarly, a combustible gas mixture of phosphine and hydrogen can be evolved through the process of water vapour induced oxidation of calcium phosphide aerosol nanoparticles. These nanoparticles can be generated as aerosol products of plasma-chemical reduction of organic calcium phosphate by organic carbon. A formation of such a small cloud of calcium phosphide nanoparticles can, for example, arise when a bird is struck by regular lightning.

Similarly, in humid air a combustible hydrogen sulphide can be generated through the process of the water vapour induced oxidation of many metal sulfide nanoparticles, for example such as natural particles of weathered iron sulfide minerals, and also some organic sulfide nanoparticles. Alternatively, a completed dry oxygen induced oxidation of metal sulfide nanoparticles can generate only noncombustible solid metal sulphates, metal oxides, as well as the sulphur dioxide and sulphur trioxide gases.

Thus, as one can see the mechanisms and products of water vapour induced oxidation of many absolutely different combustible nanoparticles radically differ from the mechanisms and products of their oxidation by oxygen gas.

Although a cone of flame is as a rule invisible over ball lightning, in the literature there are several alternative descriptions of observations of lightning balls combined with tongues of flame emerging from them [8, 9]. On the one hand, these rare observations can show that sometimes forces of a dipole-dipole attraction between separate aerosol particles-nanobatteries constituting ball lightning are inadequate to keep all the particles inside the ball cloud, and some part of the hot aerosol particles can be captured by ascending convective currents. On the other hand, these rare observations show that sometimes combustible gases, evolved during the process of water vapour induced oxidation of aerosol particles-batteries constituting ball lightning, contain components, which are able to generate visually distinguishable tongues of combustion products.

In our preliminary experiments, small burning clouds of combustible, ethanol and methanol based, droplets with the droplets diameters of ~ 1-3 micrometres were generated. A flow of these droplets suspended in the air was generated with the help of a simple ultrasonic drug inhalator with an oscillator frequency of ~ 1 megahertz. The ethanol and/or methanol based solutions contained from 2 up to 5 percent by weight of acetyl salicylic acid (99.0%, Sigma-Aldrich) plus from 2 to 3 percent by weight of calcium nitrate tetrahydrate (98.0%, Sigma-Aldrich) and plus from 2 to 5 percent by weight of cupric nitrate hemipentahydrate (98.0%, Sigma-Aldrich). It is important to note that in these experiments the micrometre-sized droplets of the combustible solutions were not exposed to additional artificial electrostatic charging. Only minimal spontaneous bipolar charging of the micrometre-sized combustible droplets took place during their ultrasonic generation.

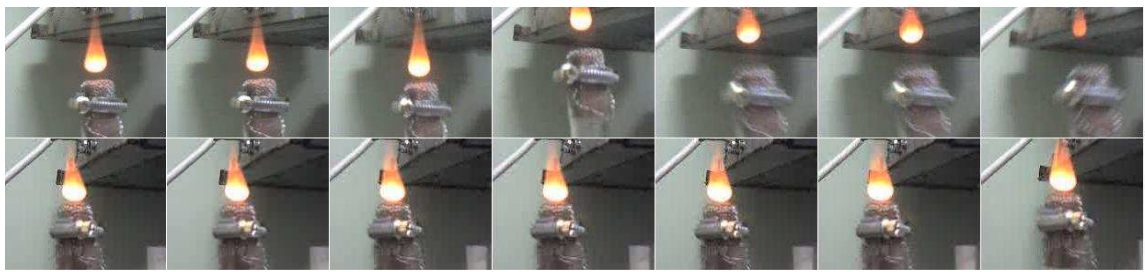

Figure 6. Practically neutral burning small clouds with a high-temperature ball-shaped core and a relatively cold and transparent cone of flame consist of billions of submicron soot based aerosol particles, which can be only minimally charged with almost balanced bipolar charges.

When burning, the droplets of the ethanol and/or methanol based solutions were subjected to immediate evaporation, initially turning into solid particles of acetyl salicylic acid mixed with calcium nitrate and copper nitrate. These solid aerosol particles in turn were exposed to a thermal oxidative decomposition to form small flame clouds consisting of submicron soot particles impregnated with a combination of calcium nitrate/calcium hydroxide/ calcium oxide and copper nitrate/ copper hydroxide/ copper oxides. Obviously, the final target products of the thermal oxidative decomposition form as aggregated soot based combustible submicrometre particles consisting of amorphous carbon mixed with the calcium and copper oxides. In addition, some other organic/inorganic solution compositions originally containing nitrates of alkaline metals also were investigated with the purpose of formation of the clouds of such carbon/air aerosol nanobatteries.

Several fragments of video of these almost electrostatically neutral burning small clouds are shown in Fig. 6. As one can see, these clouds have a high-temperature stable ball-shaped core and a relatively cold and transparent cone of flame. Diameters of the ball-shaped cores of these small burning clouds were from ~ 10 to 35 mm. These ball-shaped flame clouds could be retained as long as the aerosol flow of the micrometre-sized droplets was directed into the zone of combustion. The time of independent life of such ball-shaped small flame clouds was only of ~ 50-120 milliseconds.

It is necessary to note that video-fragments shown in Fig. 6 are used here only to experimentally illustrate the possibility of co-existence (and a visual observation) of a ball-shaped cloud (consisting of the burning hot submicron soot particles impregnated with inorganic electrolyte components) and visually distinguishable tongues of flame over this spherical cloud. Thus, the above illustration will not be connected to our further discussion, because burning particles constituting ball-shaped flame clouds shown above were only slightly charged with negligible bipolar charges, and the low bipolar charges of these aerosol particles were probably generated due to:

1) thermoionic electron emission from the surface of hot particles during combustion;

2) adsorption of negative gas ions to the surface of relatively cold aerosol particles.

The same type of thermoionic bipolar electrostatic charging of nanoparticles takes place in many processes involving dusty plasma, and probably also in the processes describing microwave induced formation of ball-shaped dusty plasma in [5, 6].

Undoubtedly, due to an intense charge-dipole attraction, in humid air both the positively and negatively charged combustible nanoparticles suspended either in a flame or in low-temperature air dusty plasma can be actively attacked and oxidized by the surrounding highly polar molecules of water vapour rather than by non-polar molecules of oxygen. However, parallel with the nanoparticle charging processes, active synchronous processes of neutralization of the bipolar charges of the flame/plasma suspended particles (caused either by cooling of the particles or by mutual recombination of opposite charges of the co-aggregated bipolar particles) can strongly reduce the final charges of these particles.

Consequently, because of gradual recombination of bipolar charges, within such practically quasi-neutral, flame or dusty plasma clouds containing overwhelming majority of uncharged or minimally charged combustible nanoparticles, the electrostatic delivery of polar molecules of water vapour to the reactive surface of these hot particles will be significantly slowed down, and so the charge-catalyzed water vapour induced oxidation of these quasi-neutral aerosol particles will be finally minimized.

Thus, if we wish to accomplish a continuous process of predominantly water vapour induced oxidation of permanently charged combustible aerosol particles, a surplus, uncompensated electrostatic charge is necessary for a cloud of such particles.

So, as assumed above and as it will be quantitatively proved later, in humid air electrostatically charged combustible particles can be exposed to preferential oxidation by surrounding polar molecules of water vapour rather than to alternative oxidation by non-polar molecules of oxygen gas due to intense charge-dipole attraction of the polar gas molecules to the charged aerosol particles. During the process of preferential water vapour induced oxidation of the charged aerosol nanoparticles, the evolving and auto-igniting of combustible gases, in particular, hydrogen gas, on the one hand, and a synchronous growing of the hydroxide based dynamic thermolabile electrolyte layers on the surface of these charged nanoparticles, on the other hand, can take place within the cloud of such nanoparticles. Both the surface growth of the hydroxide based porous electrolyte layers and evolution of the combustible reducing gas on the nanoparticles surface hamper oxidation of the nanoparticles by external neutral oxidizing species, first of all such as molecules $O_2$, but at the same time, the hydrated surface electrolyte layers can effectively transport either metal ions from reductant cores to the nanoparticles surface or negatively charged, ionized oxidizing species from the outer surface of the nanoparticles to their reductant cores, contributing to predominantly electrochemical, i.e. ion-mediated oxidation of these combustible nanoparticles. Consequently, the electrostatically charged, periodically electrohydrated combustible aerosol particles (naturally, co-aggregated with numerous surrounding polarized neutral particles become spontaneously transformed into aerosol nanobatteries periodically short-circuited by the field and thermoionic electron emission from their reductant/metal cores (the electron emitting anodes of these nanobatteries) towards external surfaces of their porous thermolabile electrolyte shells (the air cathodes of these nanobatteries) [20].

The repeating processes of auto-ignition and combustion of evolved combustible gases can make an important contribution to synchronously repeating processes of heating of the charged nanoparticles-nanobatteries and consequently to their repeating flame thermal dehydration.

During such 'thermocycling,' consisting of alternating stages of the electrostatic, charge-dipole hydration of the nanoparticles and their subsequent flame thermal dehydration, electrophysical properties of the interphase contact between a reductant, for example metal core of the nanoparticle-nanobattery and its growing, either metal oxide or metal hydroxide, shell periodically will radically change, with an alternation from the metal-semiconductor (metal-dielectric) core-shell junction (in the stage of thermal dehydration of the nanoparticle shell) to the metal-electrolyte core-shell junction (in the stage of electrostatic surface re-hydration of the temporarily cooled nanoparticle).

These fast cyclic changes of the electrophysical characteristics of the interphase core-shell contacts will result in cyclic processes of the interphase electron-ion transfer ('electron-ion jumping'), forming strong local interphase core-shell electrostatic fields, which in turn will control the specific stage-dependent mechanisms of the nanoparticles oxidation, in particular, from the Cabrera-Mott oxidation mechanism - in the stage of flame thermal surface dehydration of the combustible nanoparticle – to the electrochemical oxidation mechanism - in the stage of electrostatic surface re-hydration of this nanoparticle [29, 30, 20].

It is important to note that in the case of both the Cabrera-Mott mechanism of the nanoparticle oxidation and the electrochemical mechanism of its oxidation, each instantaneous event of gradual oxidation of combustible metal/metalloid aerosol nanoparticle will necessarily result in the generation of a powerful instantaneous local core-shell electrostatic field and consequently the generation of an instantaneous electric dipole moment of this nanoparticle.

Clearly, when a combustible aerosol particle is nano or submicrometre sized, each instantaneous event of its gas phase oxidation from one side can not be compensated by the same synchronous event of its oxidation from the other side (stochastic nature of such successive spotty oxidation of small aerosol particles is absolutely similar to a locally unbalanced character of brownian collisions). Therefore, individual nano or submicrometre combustible aerosol particles will always possess the instantaneous uncompensated electric dipole moments generated during their gas phase oxidation.

The Cabrera-Mott oxidation mechanism assumes primary migration of electrons from a metal core into a nanoporous dielectric or semi-conductor metal-oxide shell. In this case, the initial electron migration generates a local electrostatic field between the metal core and the metal oxide shell, and this field further contributes to intense outward electrodiffusion of the metal ions with their following surface oxidation.

The alternative electrochemical mechanism of oxidation assumes primary diffusion migration of metal ions from a metal core into a nanoporous surface electrolyte, for example into a more or less hydrated metal hydroxide shell. In this case, initial outward migration of the metal ions generates a local electrostatic field between the metal core and the metal hydroxide shell, and this field further contributes to intense field/ thermoionic electron emission from the metal core to the outer surface of the nanoparticle, with following surface electrochemical oxidation of the metal ions (naturally, only when these ions recombine with emitted electrons).

In fact, both these mechanisms of oxidation are electrochemical, because diffusion (or electrodiffusion) of ions through porous surface layers is a key process that precedes the events of the metal ion oxidation in both cases.

Both these types of oxidation of metal or metalloid based aerosol nanoparticles can generate powerful momentary electrostatic fields, high instantaneous electric dipole moments and strong instantaneous relaxation electron/ion core-shell currents within the nanoparticles irrespective of whether these nanoparticles are extra charged or not.

Consequently, both these mechanisms of oxidation can convert combustible aerosol nanoparticles into short-circuited nanobatteries.

It is possible to assume, however, that if a process of the metal/metalloid oxidation takes place in the real, i.e. humid air, then a local negative surface charge generated by electrons, which migrate from a metal core into growing metal oxide surface layers (according to the Cabrera-Mott oxidation mechanism) will always cause immediate electrostatic hydration of such highly charged surface sites, locally transforming these surface sites from their purely metal oxide (i.e. dielectric or semi-conductor) state into a hydrated, for example metal- hydroxide (i.e. electrolyte) state.

Thus, it seems that a process of atmospheric oxidation of many metals or metalloids can frequently be automatically converted from the Cabrera-Mott oxidation mode into the true electrochemical mode of oxidation only due to powerful fast spontaneous electrostatic hydration of oxidatively charged surface sites, which is inevitable in the conditions of the real, humid air oxidation.

It is important to emphasize that such a predominantly electrochemical mode of atmospheric oxidation of many metal structures, including metal nanoparticles, can arise only owing to natural redistribution of electrons between a metal core and a dielectric or semi-conductor metal oxide surface layer (i.e. without additional electrostatic charging of a metal object exposed to atmospheric oxidation).

Probably, in many cases, within a ball lightning cloud the heat of combustion of evolved combustible gases (for example hydrogen gas) can considerably exceed the heat that is generated or is consumed in primary heterogeneous reactions of water vapour induced oxidation of metal, or metalloid, or carbon based combustible particles constituting ball lightning.

Therefore, in particular, the highly endothermic stage of the thermal dehydration of the electrostatically charged combustible nanoparticles can be appreciably delayed in time with respect to the stage of their electrostatic oxidative hydration - only the heat of combustion of the evolved gases can effectively temporarily dehydrate the charged nanoparticles.

So, within ball lightning electrostatically charged combustible nanoparticles can be subjected to preferential oxidation by polar molecules of water vapour rather than non-polar molecules of oxygen gas. A periodic fast alternating of the processes of electrostatic hydration and thermal dehydration of the charged nanoparticles in humid air can be partially or totally synchronized with the repeating processes of evolution of combustible gases, their auto-ignition and following quick flame extinction.

Naturally, the fast alternating processes of electrostatic oxidative hydration and thermal dehydration of the charged aerosol particles constituting ball lightning can cause periodic melting and hydrogen gas induced foaming these particles; formation of submicron or micrometre-sized electrostatically charged metal/ metal oxide/ metal hydroxide hollow globules can be a natural outcome of such repeated melting/foaming.

Many witnesses who observed ball lightning at very short distances described ball lightning as a relatively low temperature object, which did not radiate intensive thermal radiation [8, 9]. Indeed a time-average temperature of the charged nanoparticles can be relatively low owing to a quick alternating of the processes of their intense heating and cooling. At the same time, peak 'colour' temperatures of the hot nanoparticles, which are periodically reached in the quickly repeating processes of their oxidative or flame re-heating, could be ~ 750-950°C (for most typical red or orange lightning balls), and so these nanoparticles could thermally emit a pulsating red or orange light with a ripple frequency high enough so that the light from ball lightning did not flicker.

Probably, however, total luminous radiation of nanoparticles-nanobatteries within ball lightning could arise from a combination of all three main light emitting processes: (1) nanoparticles-nanobatteries can emit faint light (faint thermal radiation) due to their direct oxidative heating or due to their periodic heating caused by a flame from combustion of combustible gases evolved in water vapor induced oxidative reactions; (2) nanoparticles-nanobatteries can emit pulsating luminous and ultra-violet radiation (as well as powerful pulsating wide-band radio/micro waves) at moments of their partially or totally synchronized short circuits; (3) nanoparticles-nanobatteries can probably sometimes contain (or produce during oxidation) fluorescent or phosphorescent materials (for example, some mixed metal oxides or sulphides), consequently, these particles-batteries can generate relatively low-temperature ultra-violet induced photoluminescence, also electroluminescence or cathodoluminescence.

Equally, many semi-conductor based nanoparticles-nanobatteries can probably possess properties of quantum dots. In particular, the highly charged silicon based nanoparticles - silicon/air core-shell nanobatteries [20], continuously re-transformed from the high-temperature, thermally dehydrated, Si/ $SiO_2$ core/shell nanostructure into the low-temperature, electrohydrated, Si/ $Si(OH)_4$ core/shell nanostructure, can possess properties of quantum dots [31]. Such semi-conductor based nanoparticles-nanobatteries, possessing properties of quantum dots, could be able to an intense low-temperature photo-, cathodo- and electroluminescence stimulated either by ultra-violet radiation of air-hydrogen flame or by collective discharge electron emission processes within the short-circuited nanobatteries.

Thus, a time-average temperature of many lightning balls can be relatively low, and consequently a light intensity from the ball lightning can practically be not connected with the time-average temperature of ball lightning.

**4. According to numerous witnesses' reports, iron/carbon containing objects and aluminium containing objects were frequently involved in the process of ball lightning formation**

The involvement of iron/carbon -, and/or aluminium- based objects in ball lightning formation processes was repeatedly mentioned in numerous witnesses' reports [8, 9].

In particular, a lot of these reports were devoted to descriptions of a 'high-temperature' ball lightning formation, in which a high-voltage arc evaporation of iron/carbon based materials and further condensation of the evaporated materials in the form of a small smoke cloud could be the most probable events. At the same time, alternative, relatively 'low-temperature' processes of ball lightning formation, i.e. those without the involvement of a visible electric arc, also were repeatedly described as connected with the frequent participation of the cast iron based or steel based objects [8, 9].

Probably, it would be reasonable to assume that the unipolarly charged, iron-, and/or carbon-, and/or aluminium- based combustible aerosol particles could be partially generated from the electrode materials in all such ball lightning formation phenomena.

Therefore special attention to the process of water vapour induced oxidation of the electrostatically charged iron-, or carbon-, or aluminium- based aerosol nanoparticles, or their compositions in the form of electrostatically hydrated aggregated nanobatteries (or additionally in the form of electrostatically hydrated aggregated nano-thermites) will be pertinent to our further discussion.

## 5. Some quantitative estimations of a charge-dipole interaction between a charged aerosol nanoparticle and surrounding polar gas molecules

It seems that, on the one hand, a charge-dipole interaction between a charged aerosol nanoparticle and polar gas molecules from ambient air, for example surrounding molecules of water vapour, can increase the number of collisions of these polar gas molecules with the charged nanoparticle, without a similar influence on the number of collisions of this particle with non-polar gas molecules.

On the other hand, we assume that the charge-dipole interaction between the charged aerosol nanoparticle and the surrounding polar gas molecules can in addition increase kinetic energy of the polar molecules, accelerating these gas molecules towards the charged nanoparticle at a short distance of their mean free path from the charged nanoparticle.

Let us consider a simple case of oxidation of a combustible aerosol nanoparticle charged with the minimum possible charge, i.e. charged with either a positive or negative elementary charge, $Q = |e| = 1.6 \cdot 10^{-19}$ (C).

Again, with the purpose of simplification, let us assume that:
- (a) this charged aerosol nanoparticle is spherical and it can freely and irregularly revolve on its axis due to continuous stochastic Brownian collisions with surrounding gas molecules;

(b) a time-average density of surplus electrostatic charge on the surface of this nanoparticle is practically equivalent to the time-average charge density on the surface of the same charged nanoparticle whose surplus electrostatic charge is localized in its centre, i.e. the surplus electrostatic charge is quasi-distributed on the surface of the nanoparticle, with the possibility of free fast migration of this charge on the nanoparticle surface.

In humid air, the majority of possible oxidative reactions on the surface of the discussed combustible nanoparticle can be caused by molecules of the two main competing atmospheric oxidants, oxygen gas and water vapour.

A flux of oxygen molecules incident upon the surface of the charged combustible nanoparticle suspended in humid air determines the frequency of collisions of the oxygen gas molecules with the surface of the charged nanoparticle, $f_O$ (mol/s). Correspondingly, this molecular flux determines also the rate of oxidative surface reactions caused by a fraction of the high-energy oxygen molecules possessing enough kinetic energy to climb the activation energy barriers of such oxygen induced oxidative surface reactions.

Similarly, a flux of water vapour molecules incident upon the surface of the charged combustible nanoparticle suspended in humid air determines the frequency of collisions of the water vapour molecules with the surface of the charged nanoparticle, $f_W$ (mol/s). And correspondingly, this molecular flux also determines the rate of oxidative surface reactions caused by a fraction of the high-energy water vapour molecules possessing enough kinetic energy to climb the activation energy barriers of such water vapour induced oxidative surface reactions.

## 6. Humid air oxidation of a spherical (~ 2 nm in diameter) iron metal aerosol nanoparticle charged with the minimum positive charge $Q = |e| = 1.6 \cdot 10^{-19}$ (C) at a relatively low temperature of about 300K.

As a first example, let us consider a relatively low-temperature process of humid air oxidation of a spherical (~ 2 nm in diameter) iron metal based aerosol nanoparticle charged with the minimum possible positive charge $Q = |e| = 1.6 \cdot 10^{-19}$ (C), i.e. with a single lost electron.

Let us assume that exothermic oxidation of this iron metal nanoparticle in the humid air is slow enough, because the water vapour and oxygen gas induced intense growth of the passivating layers consisting of the mixed hydrated iron hydroxides, iron oxy-hydroxides and iron oxides takes place on the surface of the iron metal nanoparticle during its humid air oxidation.

Assuming that the process of the mixed, water vapour and oxygen gas induced exothermic oxidation of the discussed charged nanoparticle is slow enough, we could also suppose, for the first example, that both a time-average temperature of the nanoparticle surface, $T_s$, and a time-average temperature of ambient humid air around this nanoparticle, $T$, can remain approximately constant and both these temperatures

are approximately equal to the initial temperature of the nanoparticle oxidation process, i.e., to ~ 300K:

$$T_s = T = 300 (K) \qquad (9)$$

First, let us compare the flux of oxygen molecules with the competing flux of water vapour molecules incident upon the surface of the discussed charged nanoparticle in the humid air.

As oxygen molecules are non-polar and their permanent electric dipole moment, $p_O$, is zero, a flux of oxygen molecules incident upon the surface of an uncharged aerosol nanoparticle is practically equal to a flux of oxygen molecules incident upon the surface of the same charged nanoparticle. In other words, the frequency of collisions of non-polar oxygen molecules with the surface of an uncharged nanoparticle, $f_{O\ unch}$ (mol/s), is practically equal to the frequency of collisions of the oxygen molecules with the surface of a similar but charged aerosol nanoparticle, $f_O$ (mol/s), which could be written as follows:

$$f_{O\ unch} = f_O = An_O v \sigma_O = An_O v \pi R^2 \qquad (10)$$

where $A$ – a numerical factor, $n_O = 0.21$ – the mole fraction of oxygen molecules in humid air (mol/mol), $v$ – the mean speed of gas molecules at a given temperature (m/s), $\sigma_O = \pi R^2$ – the cross section for collisions of surrounding non-polar oxygen molecules with the charged (or also uncharged) aerosol nanoparticle (m$^2$), $R$ – the radius of the nanoparticle (m).

If a combustible aerosol nanoparticle suspended in humid air is uncharged, non-polar molecules of atmospheric oxygen have a significant quantitative advantage over polar molecules of atmospheric water vapour when colliding (and also when reacting) with this nanoparticle, because the mole fraction of oxygen gas molecules is approximately tenfold greater than the mole fraction of water vapour molecules in humid air. Therefore non-polar oxygen molecules can be a principal gas oxidant of uncharged combustible nanoparticles in humid air, while relatively few polar molecules of water vapour probably play a secondary role in the processes of humid air oxidation of uncharged nanoparticles (Fig. 7).

The high-energy oxygen molecules possessing enough kinetic energy to climb the activation energy barriers of the oxidative surface reactions also can collide and react with the surface of uncharged combustible nanoparticle suspended in the humid air much more often in comparison with the relatively few high-energy water vapour molecules.

The preferential formation of the mixed iron oxide layers on the surface of an uncharged iron nanoparticle could be a natural outcome of such a significant quantitative advantage of oxygen gas molecules over water vapour molecules in humid air. Indeed, it seems that oxygen induced formation of layers consisting predominantly of mixed iron oxides on the surface of the uncharged iron nanoparticle in humid air is a much more probable process than an alternative process of water vapour induced formation of mixed hydrated, iron hydroxide and/or iron oxy-hydroxide, surface layers.

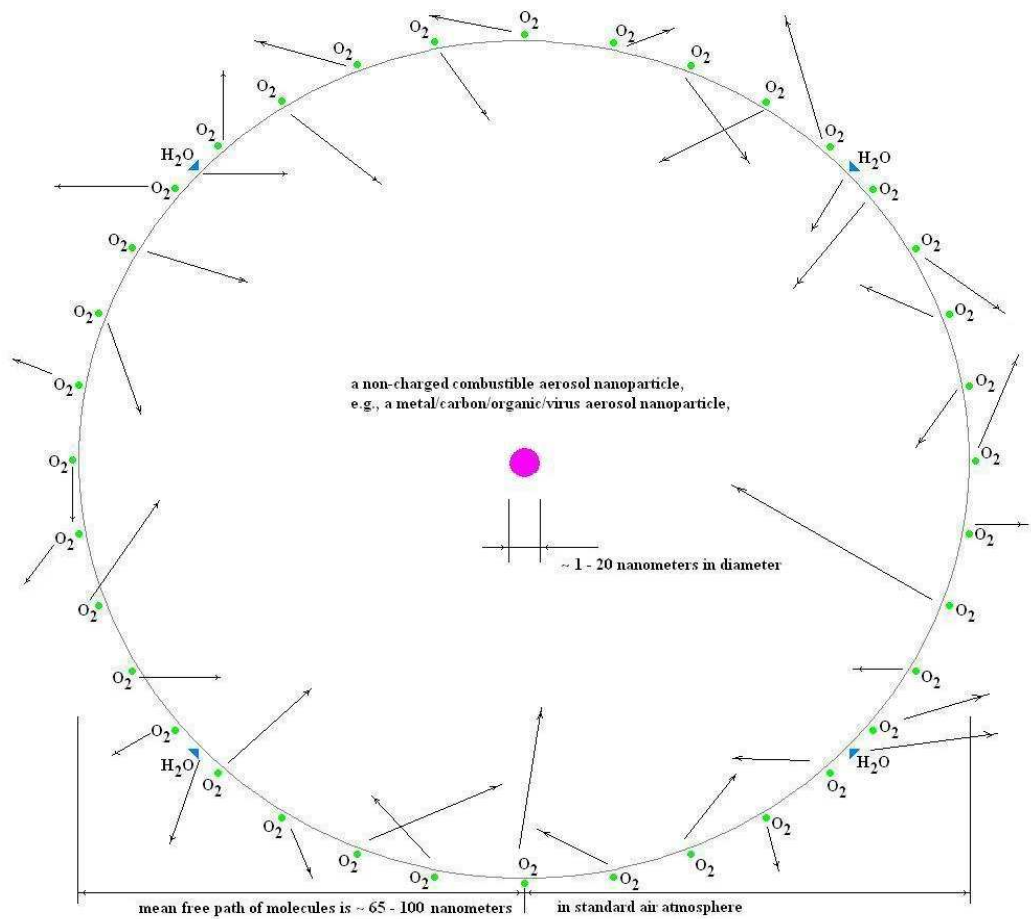

Figure 7. Even in a high humid air atmosphere, a high concentration of oxygen gas molecules creates the prerequisites for a preferential oxidation of uncharged combustible aerosol nanoparticles by oxygen molecules but not water vapour molecules. Water vapour, being another major air oxidant, with its much lower air concentration usually plays a secondary role in oxidative reactions on the surface of uncharged combustible nanoparticles. However, this situation can completely change during humid air oxidation of electrostatically charged nanoparticles.

However, probably, it would also be reasonable to suppose that intense local electrostatic hydration that converts iron oxides into iron hydroxides and iron oxy-hydroxides in humid air can still take place on oxidatively charged sites of the iron oxide surface of the originally uncharged iron nanoparticle, because electron migration from the iron metal core into the iron oxide $Fe_3O_4$ / $Fe_2O_3$ semiconductor shell will continuously generate new negative surface charges even during a predominantly oxygen gas induced oxidation of this originally uncharged iron nanoparticle.

At the same time, it seems that a different situation can arise when the discussed minimally charged iron nanoparticle is exposed to oxidation in humid air. Because water vapour molecules are high-polar, in contrast to non-polar oxygen molecules, and the permanent electric dipole moment of a water vapour molecule is

$$p_W = 1.84 \text{ (D)} = 0.6 \cdot 10^{-29} \text{ (C·m)}, \tag{11}$$

the intense charge-dipole attraction between the charged nanoparticle and surrounding polar molecules of water vapour can make an important contribution to the process of intense electrostatic nanoparticle hydration and, correspondingly, to a predominantly water vapour induced oxidation of the nanoparticle. Owing to the substantial charge-dipole attraction between the charged nanoparticle and surrounding polar molecules of water vapour, both the frequency and the intensity of collisions of the water vapour molecules with the surface of the charged aerosol nanoparticle can considerably exceed the corresponding frequency and intensity of collisions of these molecules with the surface of a similar uncharged nanoparticle (Fig. 8).

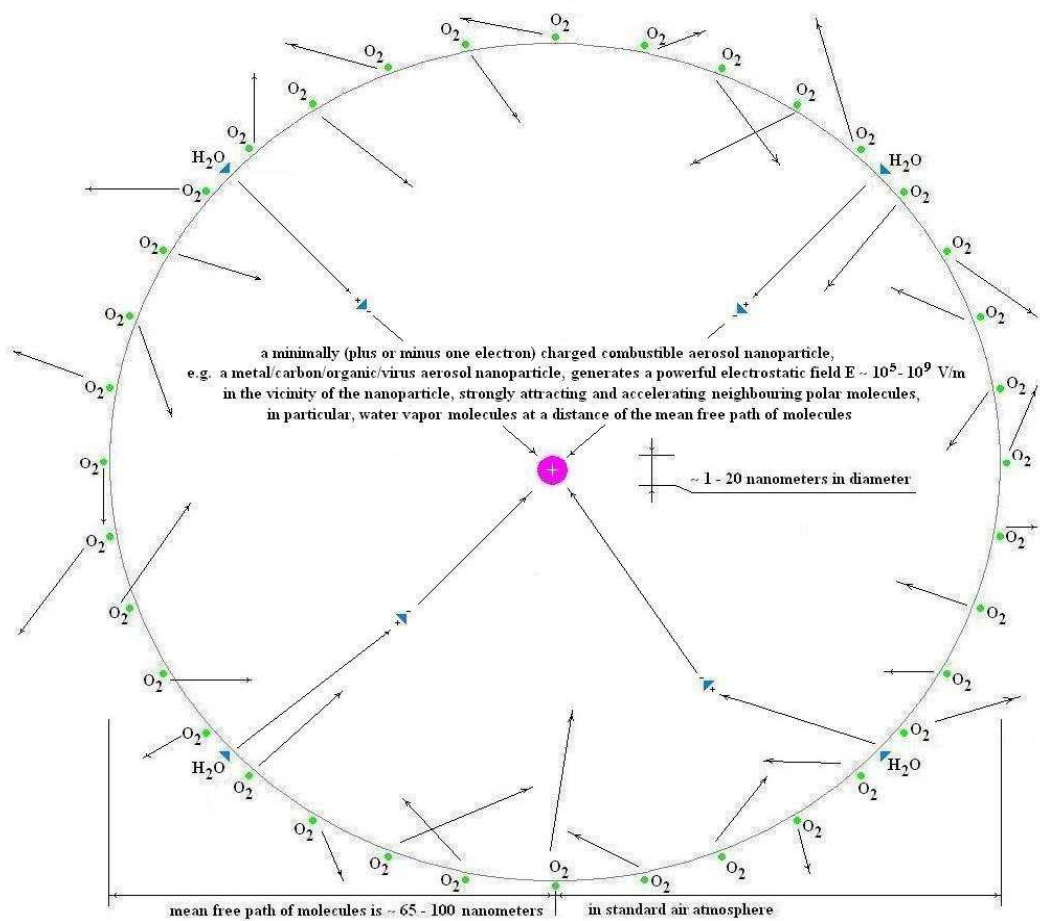

Figure 8. Powerful charge-dipole attraction between charged aerosol nanoparticles and the surrounding molecules of water vapour in humid air bends the trajectories of the polar molecules and also additionally accelerates these polar molecules in the direction of the charged nanoparticles at a distance of the air molecules' mean free path. On the other hand, non-polar air molecules, such as oxygen gas molecules (or nitrogen gas, or carbonic gas) are practically not subject to the influence of charge-dipole attraction in the vicinity of charged aerosol nanoparticles, and so the absolute values of kinetic energy and the directions of movement of the non-polar air molecules around the charged nanoparticles are practically not changed by forces of the charge-dipole attraction. A significant electrostatic acceleration of water vapour molecules in the vicinity of the charged combustible aerosol nanoparticles substantially increases reactivity of these molecules. Thus, the oxidizing efficiency of the polar molecules of water vapour in relation to the charged combustible nanoparticles considerably exceeds the alternative oxidizing efficiency of non-polar, correspondingly electrostatically non-accelerated molecules of oxygen gas, though approximately a tenfold excess of oxygen gas molecules' concentration, in comparison with the concentration of water vapour molecules, normally takes place in air atmosphere.

The frequency of collisions of polar molecules of water vapour with a surface of the charged aerosol nanoparticle, $f_W$, can be represented as follows:

$$f_W = A n_W v \sigma_W \qquad (12)$$

where $A$ – a numerical factor, $n_W \sim 0.02$ – the mole fraction of water vapour molecules in the humid atmosphere (mol/mol), $v$ – the mean speed of gas molecules at a given temperature (m/s), $\sigma_W$ – an effective cross section for collisions of surrounding polar molecules of water vapour with the charged aerosol nanoparticle (m$^2$).

The collision cross section $\sigma_W = \pi r^{*2}$ can be determined estimating the value of potential energy of the charge-dipole interaction between the charged aerosol nanoparticle and polar molecule of water vapour from a gas microenvironment of the aerosol nanoparticle (taking into account that the size of the molecular dipole, i.e. polar molecule of water vapour, is only about 0.2 nm):

$$U(r) = -\frac{1}{4\pi\varepsilon_0}\frac{pQ}{r^2} \qquad (13)$$

where

$\frac{1}{4\pi\varepsilon_0} = 9 \cdot 10^9$ (m/F), $p = p_W = 0.6 \cdot 10^{-29}$ – the permanent electric dipole moment of water vapour molecule (C·m), $Q = 1.6 \cdot 10^{-19}$ – the charge of the nanoparticle (C),

$r$ – the distance between the centre of the charged aerosol nanoparticle and the given molecule of water vapour from a gas microenvironment of the nanoparticle, (m).

An electrostatic capture of the surrounding polar molecules of water vapour with a mean kinetic energy, $W \sim 3/2\, kT$, by the aerosol nanoparticle charged with a charge, $Q$, can take place at an effective distance - the effective capture radius - $r^*$, such that

$$\frac{3}{2}kT = \frac{1}{4\pi\varepsilon_0}\frac{pQ}{r^{*2}} \qquad (14)$$

where
$k \approx 1.38 \cdot 10^{-23}$ – the Boltzmann constant (J/K),
$T$ – the temperature of ambient humid air around the nanoparticle (K).
As we assume that the temperature around the nanoparticle remains about 300 K, the mean kinetic energy of the air molecules is about:

$$W = 1.5 \cdot 1.38 \cdot 10^{-23} \text{ (J/K)} \cdot 300 \text{ (K)} = 6.21 \cdot 10^{-21} \text{ (J)} \qquad (15)$$

Correspondingly, a spherical volume of the radius of $r^*$ around the iron metal aerosol nanoparticle with an initial diameter of 2 nm, suspended in humid air and charged with a minimum positive charge $Q = |e| = 1.6 \cdot 10^{-19}$ (C), is an area where an effective electrostatic, charge-dipole capture of the surrounding polar water vapour molecules takes place at the temperature of about 300 K and at a mean atmospheric pressure at sea level, P, equal to 101,325 (Pa).

As one can see the value of the effective radius of the electrostatic capture of the surrounding water vapour molecules by the minimally charged iron nanoparticle

$$r^* = \sqrt{\frac{pQ}{4\pi\varepsilon_0 3/2kT}} = \sqrt{9 \cdot 10^9 \frac{0{,}6 \cdot 10^{-29} \cdot 1{,}6 \cdot 10^{-19}}{3/2 \cdot 4{,}1 \cdot 10^{-21}}} = 1.19 \cdot 10^{-9} \text{ (m)} \quad (16)$$

only slightly exceeds the 1 nm radius of the discussed charged nanoparticle and also it is much less than the mean free path of the ambient gas molecules, $l \approx 10^{-7}$(m), under the given conditions of temperature and pressure.

The corresponding cross section for collisions of the surrounding polar molecules of water vapour with the discussed charged iron nanoparticle (and probably also for electrostatic capture of these polar molecules by the charged iron nanoparticle) is

$$\sigma_W \sim \pi r^{*2} = 4.5 \cdot 10^{-18} \text{ (m}^2\text{)} \quad (17)$$

This cross section also only slightly (approximately one and a half times) exceeds a 'geometrical' cross section for possible collisions of this charged nanoparticle with the surrounding non-polar oxygen molecules:

$$\sigma_O \sim 3.14 \cdot 10^{-18} \text{ (m}^2\text{)} \quad (18)$$

Indeed as one can see the cross section for collisions of the surrounding polar molecules of water vapour with a minimally charged 2 nm diameter aerosol nanoparticle can only slightly exceed the cross section for collisions of this aerosol nanoparticle with the surrounding non-polar molecules of oxygen gas due to a contribution of the charge-dipole attraction between such a charged aerosol nanoparticle and surrounding polar molecules of water vapour.

However, the situation strongly changes when the diameter of the minimally charged nanoparticle becomes less than 1 nm. For example, it is easy to estimate that a cross section for collisions of a minimally charged 1 nm diameter aerosol nanoparticle with the surrounding water vapour molecules can approximately by sixfold exceed the cross section for collisions of a similar minimally charged aerosol nanoparticle with the surrounding non-polar gas molecules, including oxygen gas molecules.

Correspondingly, in a humid atmosphere, electrostatic hydration of such small charged aerosol nanoparticles (or such small substrate-integrated charged nanoscale surface sites) can be much more intense than alternative hydration of similar uncharged nanoparticles (or uncharged surface sites of substrate-integrated nanostructures).

In fact, minimally spontaneously charged aerosol nanoparticles (equally minimally charged substrate-integrated nanostructures) with characteristic sizes of ≤ 1 nm are extremely widespread nanotechnological and, particularly, biological objects.

A good example of one of the useful nanotechnological applications of powerful electrostatic hydration of such minimally charged substrate-integrated objects, whose diameters are ≤ 1 nm, is the well-known water-assisted growth of carbon nanotubes, so-called the super-growth synthesis of carbon nanotubes [32-33]. During such water-assisted carbon nanotube's growth, an electric contact between the growing single-walled carbon nanotubes and their substrate causes a natural contact charging of these nanotubes, whose diameters and wall thickness are ≤ 1 nm. Corresponding intense electrostatic hydration of the unipolarly charged ends of the carbon nanotubes will cause a reduction of mutual electrostatic repulsion of these highly charged ends and provoke dense surface packing of voids within the forest of the growing carbon nanotubes by electrostatically adsorbed water vapour molecules. In addition, this intense electrostatic hydration of the ends of the contact charged carbon nanotubes can effect the continuous oxidative re-uncapping of these ends by the surrounding electro-accelerated molecules of water vapour, which probably can also promote the intense prolonged ordered growth of the carbon nanotubes.

In a humid atmosphere, electrostatical hydration of charged aerosol nanoparticles (or charged substrate-integrated nanostructures) can be particularly intense within a very small charged area with a radius of ≤ 1 nm, where the local charge (e.g. an adsorbed hydrated ion), is located at the centre of this area that is most actively bombarded by the surrounding polar molecules, locally electrostatically accelerated.

Because this zone with a highest possible charge density and with a diameter of ≤ 2 nm, exposed to the most active attack by surrounding polar molecules, is able to migrate (together with its charge) onto the surface of an individual combustible aerosol particle or along chains consisting of aggregated aerosol particles, then the processes of charge-catalyzed water vapour induced oxidation can involve a high number of initially uncharged particles.

However, let us return to the question about humid air oxidation of the discussed minimally charged spherical iron nanoparticle with a diameter of ~ 2 nm.

Despite the fact that $\sigma_W \approx 1.5\sigma_O$, when comparing oxygen and water vapour molecular fluxes incident upon the surface of the discussed iron nanoparticle, one can see that the frequency of collisions of the surrounding polar molecules of water vapour with the surface of this nanoparticle, $f_W = An_W v \sigma_W$, is still much lower than the frequency, $f_O = An_O v \sigma_O$, of alternative collisions of the surrounding oxygen molecules with this surface:

$$f_W < f_O \qquad (19)$$

because the mole fraction of atmospheric molecules of oxygen gas $n_O = 0.21$ is much higher than the mole fraction of water vapour molecules $n_W \sim 0.02$ in humid air.

It is clear that this result does not depend on whether the discussed combustible nanoparticle is charged positively, with a single lost electron, or if it is charged negatively, with a single surplus electron.

Thus, we can draw a conclusion that in humid air the charge-dipole attraction between the minimally charged aerosol nanoparticle with a diameter 2 nm or greater and the surrounding polar molecules of water vapour does not provide a noticeable quantitative advantage of polar molecules of water vapour over non-polar molecules of oxygen in access to a surface of this minimally charged nanoparticle.

However, there is another important aspect to be discussed, namely, the influence of the charge-dipole attraction between the charged aerosol nanoparticle and the surrounding polar molecules of water vapour on the intensity of collisions of these polar molecules with the surface of the nanoparticle.

Indeed, in humid air the charge-dipole attraction between the discussed charged combustible nanoparticle and the surrounding polar molecules of water vapour are not able to appreciably increase the frequency of collisions of the polar molecules with such a large minimally charged nanoparticle.

But the charge-dipole attraction in addition can substantially accelerate the surrounding polar molecules of water vapour in the direction of the nanoparticle at the distance of the mean free path of ambient gas molecules $l \approx 10^{-7}$(m) under the given conditions of temperature and pressure, increasing the number of high-energy molecules of water vapour with enough kinetic energy to climb the activation energy barriers of the oxidative reactions induced by water vapour on the surface of the charged combustible aerosol nanoparticle.

Let us now estimate this possibility in order to compare the rates of the corresponding oxidative reactions potentially induced on the surface of the charged iron nanoparticle by either non-polar high-energy molecules of oxygen or electrostatically extra accelerated polar high-energy molecules of water vapour.

## 7. Additional average kinetic energy acquired by polar molecules of water vapour in an electrostatic field of a minimally charged aerosol nanoparticle at the mean free path distance from the surface of this nanoparticle

Additional average kinetic energy $\varepsilon_1$ acquired by the polar molecule of water vapour in the electrostatic field of the iron nanoparticle with an initial diameter of 2 nm, suspended in humid air and charged with the minimum positive charge $Q = |e| = 1.6 \cdot 10^{-19}$ (C), at an acceleration distance of the mean free path from the nanoparticle is

$$\varepsilon_1 = \frac{1}{4\pi\varepsilon_0}\frac{pQ}{R^2} - \frac{1}{4\pi\varepsilon_0}\frac{pQ}{l^2} \approx \frac{1}{4\pi\varepsilon_0}\frac{pQ}{R^2} \qquad (20)$$

where

$\frac{1}{4\pi\varepsilon_0} = 9 \cdot 10^9$ (m/F), $p = p_W = 0.6 \cdot 10^{-29}$ – the permanent electric dipole moment of a water vapour molecule (C·m), $Q = 1.6 \cdot 10^{-19}$ – the charge of the nanoparticle (C), $R = 1.0 \cdot 10^{-9}$ – the radius of the nanoparticle (m), $l \approx 1.0 \cdot 10^{-7}$ – the mean free path of ambient gas molecules (m).

Thus, the additional average kinetic energy $\varepsilon_1$ acquired by the polar molecule of water vapour in the electrostatic field of the discussed minimally charged nanoparticle at a distance of the mean free path from the nanoparticle is

$$\varepsilon_1 = 9 \cdot 10^9 \cdot 0.6 \cdot 10^{-29} \cdot 1.6 \cdot 10^{-19}/10^{-18} \approx 8.6 \cdot 10^{-21} \text{ (J)} \qquad (21)$$

The additional average energy of electrostatic acceleration of polar molecules of water vapour appreciably, more than twice (!), exceeds the characteristic thermal energy, $kT$, of the surrounding air molecules at a temperature of about 300K:

$$kT = 4.1 \cdot 10^{-21} \text{ (J)} \tag{22}$$

According to [21], "the activation energy for the reaction of Fe with O2 was determined to be 32 ± 6 kJ/mol for exposure times of 20–2000s and 28 ± 3 kJ/mol for the reaction of Fe with water vapour for exposure times of 100–2000s". Correspondingly, we will suppose that the activation energy for the process of water vapour induced oxidation of iron metal nanoparticle, $E_{aW}$ (J/mol), also is about 28,000 (J/mol) at a temperature of about 300K. And correspondingly, the activation energy for an alternative competing reaction of oxygen gas induced oxidation of iron metal nanoparticle, $E_{aO}$ (J/mol), also is about 32,000 (J/mol) at the given temperature.

These, practically equivalent activation energies, $E_{aW} \approx E_{aO} = E_a \approx 30,000$ (J/mol), can be calculated per one molecule, $\varepsilon_a$ (J), and, correspondingly, are about

$$\varepsilon_a = E_a/N_A \approx 30,000/(6.02 \cdot 10^{23}) \approx 5 \cdot 10^{-20} \text{ (J)} \tag{23}$$

where $N_A \approx 6.02 \cdot 10^{23}$ – the Avogadro constant (mol$^{-1}$).

One can see that the ratio of the activation energy, $\varepsilon_a$ (J), for both reactions of iron oxidation (with the participation of either the surrounding water vapour or oxygen gas molecules) to the characteristic thermal energy, $kT$, of these air molecules at the temperature of about 300K is

$$\varepsilon_a/kT = 5 \cdot 10^{-20} / 4.1 \cdot 10^{-21} = 12.2 \tag{24}$$

Only high energy molecules of water vapour and oxygen gas, with kinetic energy $\varepsilon \geq \varepsilon_a$, can react with the discussed minimally charged aerosol iron nanoparticle.

The number of such reactive molecules is equal to:

$$\Delta N = N \frac{2}{\sqrt{\pi}} \frac{1}{(kT)^{3/2}} \int_{\varepsilon_a}^{\infty} \sqrt{\varepsilon} e^{-\varepsilon/(kT)} d\varepsilon = N \frac{2}{\sqrt{\pi}} \int_{\varepsilon_a/(kT)}^{\infty} \sqrt{\varepsilon'} e^{-\varepsilon'} d\varepsilon'. \tag{25}$$

By integration by parts and a series expansion, we can approximately calculate a relative quantity of reactive oxygen gas molecules possessing enough kinetic energy to climb the activation energy barrier of the oxygen induced oxidative reactions on the surface of the discussed charged iron nanoparticle at a temperature of about 300K:

$$\Delta N_O / N_O = \frac{2}{\sqrt{\pi}} \int_{12.2}^{\infty} \sqrt{\varepsilon'} e^{-\varepsilon'} d\varepsilon' \approx \frac{2}{\sqrt{\pi}} e^{-12.2} \left( \sqrt{12.2} + \frac{1}{2\sqrt{12.2}} \right) = 2.1 \cdot 10^{-5} \tag{26}$$

Clearly, to calculate a relative quantity (a fraction) of reactive water vapour molecules possessing enough kinetic energy to climb the activation energy barrier of the water vapour induced oxidative reactions on the surface of the discussed charged iron nanoparticle at the temperature of about 300K, it is necessary to integrate the above expression with the lower limit of integration:

$$\varepsilon_a / (kT + \varepsilon_1) = 5 \cdot 10^{-20} / (4.1 \cdot 10^{-21} + 8.6 \cdot 10^{-21}) = 3.9. \tag{27}$$

Let us recall that $\varepsilon_1 \approx 8.6 \cdot 10^{-21}$ (J) is the additional average kinetic energy acquired by the polar molecule of water vapour in the electrostatic field of minimally charged nanoparticle at a distance of the mean free path from the nanoparticle.

Correspondingly, the fraction of the reactive water vapour molecules possessing enough kinetic energy to climb the activation energy barrier of the water vapour induced oxidative reactions on the surface of the discussed charged iron nanoparticle at a temperature of about 300K is:

$$\Delta N_H / N_H = \frac{2}{\sqrt{\pi}} \int_{3.9}^{\infty} \sqrt{\varepsilon'} e^{-\varepsilon'} d\varepsilon' \approx \frac{2}{\sqrt{\pi}} e^{-3.9} \left( \sqrt{3.9} + \frac{1}{2\sqrt{3.9}} \right) = 0.049 \tag{28}$$

As one can see, this fraction of the additionally electrostatically accelerated reactive water vapour molecules, $\Delta N_H / N_H = 4.9 \cdot 10^{-2}$, substantially exceeds the alternative fraction of the reactive oxygen gas molecules $\Delta N_O / N_O = 2.1 \cdot 10^{-5}$.

Recalculating these relative quantities of oxygen and water vapour reactive molecules, which are able to climb the activation energy barriers of the oxidative reactions on the surface of the discussed charged iron nanoparticle at a temperature of about 300K, per one mole of humid air, and taking into account the concentration difference (typical for humid air) between these main air oxidants, we have

$$N_O = N_A \cdot n_O = 6.02 \cdot 10^{23} \cdot 0.21 = 1.26 \cdot 10^{23} \tag{29}$$

where $N_O$ – the quantity of oxygen molecules in one mole of humid air,

$N_A \approx 6.02 \cdot 10^{23}$ – the Avogadro constant (mol$^{-1}$), $n_O = 0.21$ – the mole fraction of oxygen molecules in humid air (mol/mol).

Analogously,

$$N_W = N_A \cdot n_W = 6.02 \cdot 10^{23} \cdot 0.02 = 1.2 \cdot 10^{22} \tag{30}$$

where

$N_W$ – the quantity of water vapour molecules in one mole of humid air,

$N_A \approx 6.02 \cdot 10^{23}$ – the Avogadro constant (mol$^{-1}$), $n_W \sim 0.02$ – the mole fraction of water vapour molecules in humid air (mol/mol), which could for example correspond to normal summer atmospheric conditions with a relative humidity of ~ 80 % and a temperature of ~ 300K (i.e. ~ 27°C).

Now, we can calculate:

(a) the absolute quantity of reactive oxygen molecules, which are able to climb the activation energy barrier of the oxidative reactions on the surface of the discussed minimally charged iron nanoparticle at a temperature of about 300K, and which are contained in one mole of humid air:

$$\Delta N_O = N_O \cdot 2.1 \cdot 10^{-5} = 1.26 \cdot 10^{23} \cdot 2.1 \cdot 10^{-5} = 2.6 \cdot 10^{18} \qquad (31)$$

(b) the absolute quantity of reactive molecules of water vapour, which are able to climb the activation energy barrier of the oxidative reactions on the surface of the discussed minimally charged iron nanoparticle at a temperature of about 300K, and which are contained in one mole of humid air:

$$\Delta N_H = N_H \cdot 0.049 = 1.2 \cdot 10^{22} \cdot 0.049 = 588 \cdot 10^{18} \qquad (32)$$

As one can see, the number of reactive water vapour molecules, which are able to oxidize the discussed minimally charged iron nanoparticle suspended in humid air at a temperature of ~ 300K is

$$\Delta N_H / \Delta N_O = 588 \cdot 10^{18} / 2.6 \cdot 10^{18} = 226 \qquad (33)$$

times larger than the number of reactive oxygen molecules, which are able to oxidize the charged surface of this nanoparticle under similar atmospheric conditions when the mole fraction of water vapour molecules, $n_W$, in humid air is ~ 0.02.

If the mole fraction of water vapour molecules, $n_W$, in humid air is only 0.01 (which could for example correspond to atmospheric conditions with a relative humidity of ~ 57 % and an air temperature of ~ 293K, i.e. ~ 20°C, or which could equally correspond to atmospheric conditions with a relative humidity of ~ 100 % and an air temperature of ~ 282K, i.e. ~ 9°C), even in these cases the number of water vapour molecules, which are able to oxidize the discussed minimally charged aerosol iron nanoparticle is only

$$\Delta N_H / \Delta N_O = 226 / 2 = 113 \qquad (34)$$

times larger than the number of oxygen molecules, which are able to oxidize this charged nanoparticle under such conditions.

One can see that despite the relatively small number of polar molecules of water vapour in normal humid air, they can become a major air oxidant of charged aerosol nanoparticles (or, analogously, charged surface nanostructures) due to the influence of the charge-dipole interaction between these charged aerosol nanoparticles (charged surface nanostructures) and surrounding polar molecules of water vapour.

Indeed, it is interesting that the above calculated number of reactive molecules of water vapour, which are able to oxidize charged iron nanoparticles can hundreds of times exceed the number of reactive molecules of oxygen gas, competitively attacking these charged nanoparticles at an air temperature of about 300K in normal conditions of widely varying air humidity.

Although this effect of preferential oxidation of charged combustible nanoparticles by water vapour molecules rather than by oxygen gas molecules contained in humid air is shown for a possible process of oxidation of charged iron metal nanoparticles, it seems that the similar effect can take place during humid air oxidation of a lot of various, not only metal based combustible nanoparticles.

The effect of electrostatic, charge-dipole acceleration of polar gas molecules in immediate proximity to a charged aerosol nanoparticle does not depend on the polarity of the nanoparticle charge. Thus, both the positively charged nanoparticles and negatively charged nanoparticles equally can electrostatically attract and accelerate the surrounding polar gas phase molecules, including water vapour molecules, at the distance of the mean free path of these molecules from the charged nanoparticles.

The effect of the electrostatic acceleration of polar molecules near to the charged nanoparticle also does not depend on the specific material components of the charged nanoparticle.

During oxidation, a final diameter of the completely oxidized nanoparticle, originally consisting of iron metal, can be only slightly increased in comparison with the initial diameter of the iron metal nanoparticle due to the oxidative growth of the iron hydroxide and/or iron oxy-hydroxide based porous surface layers, which are less dense than iron metal. However, it is simple to estimate that a change in the diameter of the iron metal nanoparticle during this oxidation process does not exceed 50% of the initial diameter of the iron metal nanoparticle.

Now, let us consider the analogous process of humid air oxidation of the spherical iron metal aerosol nanoparticle with a diameter of 2 nm charged with the minimum positive charge $Q = |e| = 1.6 \cdot 10^{-19}$ (C) at the higher temperature of about 573K.

## 8. Humid air oxidation of a spherical (~ 2 nm in diameter) iron metal aerosol nanoparticle charged with the minimum positive charge $Q = |e| = 1.6 \cdot 10^{-19}$ (C) at a relatively high temperature of about 573K

Let us suppose again that in humid air the mole fraction of molecules of water vapour, $n_W$, is ~ 0.02.

Let us suppose also that at a temperature of ~ 573 K (i.e. ~ 300°C), for the competing processes of water vapour and oxygen gas induced oxidation of iron metal surface, i.e. for the oxidative processes with high temperature parabolic kinetics, an average value, $E_a$ (J/mol), of both activation energies [i.e. the activation energy for water vapour induced oxidation of iron nanoparticle, $E_{aW}$ (J/mol), and the activation energy for oxygen gas induced oxidation of this nanoparticle $E_{aO}$ (J/mol)] can be of ~ 150,000 (J/mol) [21, 34-36]:

$$E_a \approx E_{aW} \approx E_{aO} \sim 150{,}000 \text{ (J/mol)} \qquad (35)$$

An essential deceleration of the oxidative processes on the surface of the iron metal nanoparticle takes place due to a passivating influence of the growing surface layers, which probably consist mainly of mixed iron hydroxide $Fe(OH)_2$ and iron oxy-hydroxide $FeOOH$ (at this temperature), and which are exposed to continuous saturation with hydrogen gas evolved on and within the growing surface layers [22] due to reaction between iron metal nanoparticle and electrostatically absorbed water vapour. Thus, the growing iron hydroxide based surface layers and evolving hydrogen gas can probably significantly inhibit iron metal oxidation in humid air (Fig. 9).

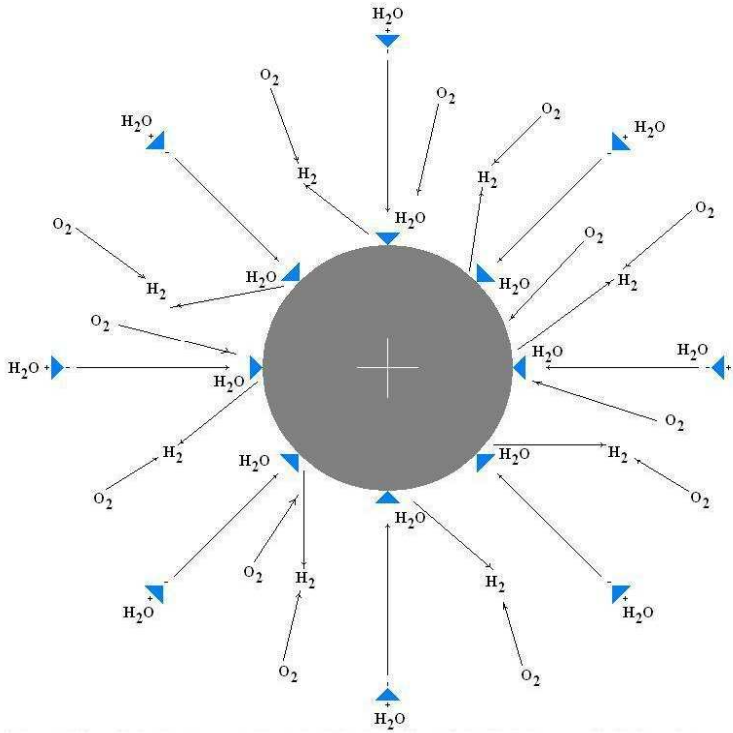

Figure 9. An attack of surrounding molecules of water vapour against charged combustible aerosol nanoparticles is much more intense in comparison with an alternative attack of non-polar molecules of oxygen gas, because a powerful charge-dipole acceleration of the polar molecules of water vapour towards the charged nanoparticles takes place in immediate proximity to the nanoparticles. In such cases, at an ambient air temperature of ~ 300 K the only elementary charge, either a positive or negative one, when located on the surface of the nanoparticle, can significantly (approximately twice) increase the kinetic energy of the collisions of the surrounding water vapour molecules with the surface of this charged nanoparticle due to the powerful charge-dipole attraction of the water vapour molecules to the charged surface or, more precisely, due to a powerful electrostatic attraction of surrounding water vapour molecular dipoles into regions on the surface of the charged nanoparticle with a maximum charge density and a corresponding maximum local electrostatic intensity. Even at ambient air temperatures of ~ 500-700K, the water vapor molecules, electrostatically accelerated around the highly charged aerosol nanoparticles, can collide with the surface of the nanoparticles with kinetic energy still greater than the average kinetic energy of the rest of the air molecules in a thermal equilibrium. This means that accelerated polar molecules of water vapour can effectively oxidize the charged combustible nanoparticles in humid air under such temperature conditions when thermally equilibrium non-polar molecules, including oxygen gas molecules, remain practically inactive. In humid air, molecules of water vapour can become a major oxidant in relation to various highly charged combustible nanoparticles, in both their aerosol and/or precipitated form.

Thus, at $T \approx 573$ K:

$$kT \approx 1.38 \cdot 10^{-23} \cdot 573 = 7.91 \cdot 10^{-21} \text{ (J)} \tag{36}$$

$$\varepsilon_a = E_a / N_A \approx 150{,}000 / (6.02 \cdot 10^{23}) \approx 2.5 \cdot 10^{-19} \text{ (J)} \tag{37}$$

where $N_A \approx 6.02 \cdot 10^{23}$ – the Avogadro constant (mol$^{-1}$).

Consequently, in the case of this relatively high temperature humid air oxidation of the charged iron metal aerosol nanoparticle:

$$\varepsilon_a/kT = 31.6 \tag{38}$$

$$\varepsilon_a/(kT + \varepsilon_1) = 2.5 \cdot 10^{-19} / (7.91 \cdot 10^{-21} + 8.6 \cdot 10^{-21}) = 15.2 \tag{39}$$

Let us again recall that $\varepsilon_1 \approx 8.6 \cdot 10^{-21}$ (J) is the additional average kinetic energy acquired by the polar molecule of water vapour in the electrostatic field of minimally charged nanoparticle at a distance of the mean free path from the nanoparticle.

Thus, the fraction of the reactive oxygen gas molecules, i.e. oxygen gas molecules possessing enough kinetic energy to climb the activation energy barrier of the oxygen gas induced oxidative reactions on the surface of the discussed iron nanoparticle at a temperature of about 573K, is:

$$\Delta N_O / N_O = \frac{2}{\sqrt{\pi}} \int_{31.6}^{\infty} \sqrt{\varepsilon'} e^{-\varepsilon'} d\varepsilon' \approx \frac{2}{\sqrt{\pi}} e^{-31.6} \left( \sqrt{31.6} + \frac{1}{2\sqrt{31.6}} \right) = 1.22 \cdot 10^{-13} \tag{40}$$

Consequently, the competing fraction of the reactive water vapour molecules possessing enough kinetic energy to climb the activation energy barrier of the water vapour induced oxidative reactions on the surface of the discussed minimally charged iron nanoparticle at a temperature of about 573K is:

$$\Delta N_H / N_H = \frac{2}{\sqrt{\pi}} \int_{15.2}^{\infty} \sqrt{\varepsilon'} e^{-\varepsilon'} d\varepsilon' \approx \frac{2}{\sqrt{\pi}} e^{-15.2} \left( \sqrt{15.2} + \frac{1}{2\sqrt{15.2}} \right) = 1.14 \cdot 10^{-6} \tag{41}$$

Thus, in this relatively high temperature case of oxidation of the charged iron metal particle:

(a) the absolute quantity of the reactive non-polar molecules of oxygen gas, which is able to climb the activation energy barrier of the oxidative reactions on the surface of the discussed minimally charged nanoparticle at a temperature of about 573K, and which are contained in one mole of humid air:

$$\Delta N_O = N_O \cdot 1.22 \cdot 10^{-13} = 1.26 \cdot 10^{23} \cdot 1.22 \cdot 10^{-13} = 1.53 \cdot 10^{10} \tag{42}$$

(b) the absolute quantity of the reactive polar molecules of water vapour, which is able to climb the activation energy barrier of the oxidative reactions on the surface of the discussed minimally charged nanoparticle at a temperature of about 573K, and which are contained in one mole of humid air:

$$\Delta N_H = N_H \cdot 1.14 \cdot 10^{-6} = 1.2 \cdot 10^{22} \cdot 1.14 \cdot 10^{-6} = 1.37 \cdot 10^{16} \tag{43}$$

As one can see, the number of reactive water vapour molecules, which are able to oxidize the discussed minimally charged iron nanoparticle suspended in humid air at a temperature of ~ 573K is

$$\Delta N_H / \Delta N_O = 1.37 \cdot 10^{16} / 1.53 \cdot 10^{10} = 0.9 \cdot 10^6 \tag{44}$$

times larger than the number of reactive oxygen gas molecules, which are able to oxidize the charged surface of this nanoparticle under such relatively high temperature conditions of humid air oxidation when the mole fraction of water vapour molecules, $n_W$, in humid air is ~ 0.02.

Indeed, an unexpectedly large (approximately one million times) difference is found between the number of reactive water vapour molecules, which are able to oxidize the minimally charged iron nanoparticle in humid air and the number of alternative reactive molecules of oxygen gas, competitively participating in this humid air oxidation. This means that relatively few polar molecules of water vapour in humid air, being additionally electrostatically accelerated by their charge-dipole attraction towards the minimally charged iron nanoparticle, become practically the only air oxidant of the discussed minimally charged iron nanoparticle, and non-polar molecules of oxygen gas actually do not take part in humid air oxidation of this charged nanoparticle both at an oxidation temperature of ~ 300K and especially at an oxidation temperature of ~ 573K.

Thus, probably, that one of the main oxidative reactions on the surface of the minimally charged iron nanoparticle will be a water vapour induced exothermal reaction of oxidation of iron metal with formation of the hydrated, iron hydroxide $Fe(OH)_2$ and/or iron oxy-hydroxide $FeOOH$ based porous electrolyte surface layers with synchronous evolution of hydrogen gas:

$$3\ Fe_{(nanoparticle)} + 6\ H_2O_{(vapour)} = Fe(OH)_2 + 2FeOOH + 4\ H_2\uparrow \qquad (45)$$

The solid reaction products, growing on the surface of the charged iron nanoparticle in the form of the hydrated electrolyte layers, will limit diffusion and direct oxidation of the nanoparticle by external neutral oxidizing species, but at the same time, these electrolyte layers will effectively transport:

(a) ionized oxidizing species, such as $OH^-$ ions and $O_2^-$ ions into the iron core;
(b) iron metal ions, such as $Fe^{2+}$, or $Fe^{3+}$, or $Fe(OH)^+$, $Fe(OH)^{2+}$, in the opposite direction, to the outer surface of the iron metal nanoparticle.

The solid reaction products in the form of the hydrated hydroxide or oxy-hydroxide surface electrolyte layers are quite thermostable, and reactions of their thermal decomposition proceed only at relatively high temperatures:

$$Fe(OH)_2 = FeO + H_2O_{(vapour)} \qquad 423 - 473K \qquad (46)$$

$$2\ FeOOH = Fe_2O_3 + H_2O_{(vapour)} \qquad 773 - 973K \qquad (47)$$

$$2\ FeOOH + Fe(OH)_2 = Fe_3O_4 + 2\ H_2O_{(vapour)} \qquad 873 - 1273K \qquad (48)$$

Therefore electrochemical oxidation of the discussed minimally charged iron metal nanoparticle is the process that can effectively compete with the process of oxidation of this nanoparticle by neutral oxidizing species even at quite high temperatures.

At such temperatures, hydrogen gas evolved in exothermal reaction (45), can be auto-ignited directly on the surface of the discussed aerosol nanoparticle in humid air, because an autoignition temperature of the air-hydrogen gas mixtures is about 858K (i.e., 585° C).

Preferential electrochemical oxidation of the continuously electrohydrated charged iron metal nanoparticles in humid air will convert these nanoparticles into the short-circuited iron/air core-shell nanobatteries, or more precisely into the short-circuited iron metal/ iron oxy-hydroxide/ water vapour nanobatteries, i.e. into the short-circuited core-shell nanobatteries with the iron metal based core reductant, with the iron hydroxide/iron oxy-hydroxide based porous surface electrolyte shell, and with the external gas oxidant – i.e. predominantly with water vapour (Fig. 10).

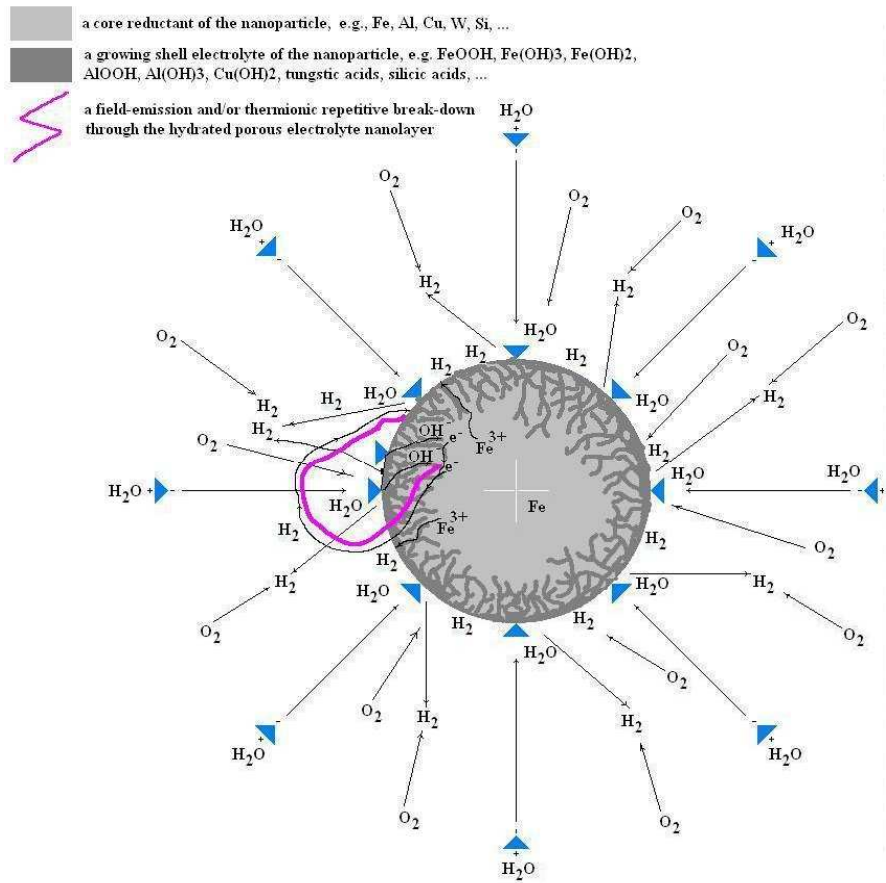

Figure 10. The electrostatic charge of combustible nanoparticles can be a powerful catalyst of their intense oxidation by polar molecules of water vapour in a humid atmosphere. Due to the additional energy of electrostatic acceleration, polar molecules of water vapour can actively oxidize various highly charged nanoparticles, including nanoparticles consisting of materials, which are relatively inert at temperatures of ≤ 700K (for example, such as soot and other carbon based nanoparticles that, when uncharged, are inert in humid air at these temperatures).
During the process of such charge-catalyzed humid air oxidation of combustible nanoparticles by surrounding water vapour molecules, various combustible gases can be generated (e.g. hydrogen gas when oxidizing nanoparticles, consisting of some reactive metals; or a mixture of hydrogen gas and carbon monoxide when oxidizing some carbon-containing nanoparticles; or hydrogen sulphide gas when oxidizing some metal sulphide nanoparticles).
Repeating processes of auto-ignition of evolved combustible gases can accompany water vapour induced oxidation of charged combustible nanoparticles in humid air. When the auto-ignition of the combustible gases takes place within such a cloud, a flame is not always visible. In the case of the water vapour induced oxidation of charged metal/ metalloid nanoparticles, e.g. the iron or aluminium or silicon ones, the evolved hydrogen gas can be auto-ignited without a visible flame.

Thus, the charged iron metal nanoparticle can be spontaneously transformed into the nanoscale battery with an internal reductant - iron metal core - an anode of the battery.

During the process of humid air oxidation, the internal iron metal anode is periodically emitting electrons outwards, through the porous surface electrolyte layers, to the outer surface of the nanoparticle, i.e. to a virtual cathode of this core-shell nano capsule-nanobattery.

The diffusion flux of the oxidizable iron-containing ions, i.e. $Fe^{2+}$, or $Fe^{3+}$, or $Fe(OH)^+$, or $Fe(OH)^{2+}$, also moves from the iron metal core through the FeOOH/Fe(OH)$_2$ containing surface electrolyte layers to the outer surface of the charged iron metal nanoparticle. These positively charged iron-containing ions are exposed to oxidation on and within the hydroxide based nanoparticle shell where these ions meet and recombine with electrons that are periodically emitted from iron metal core (i.e. only when the electrochemically generated intra-particle electrostatic intensity reaches limiting values permitting field and/or thermionic electron emission from the iron metal core.

One of numerous possible oxidative reactions, which can take place directly on the hydrated iron hydroxide based surface of the charged iron metal nanoparticle-nanobattery, with the participation of the surrounding electrostatically accelerated water vapour molecules and the oxidizable iron-containing ions:

$$Fe(OH)^{2+}_{(ion)} + 2e + 2H_2O_{(vapour)} \rightarrow Fe(OH)_3 + H_2 \qquad (49)$$

The counter diffusion flux of negative hydroxyl ions through the FeOOH/Fe(OH)$_2$ containing surface electrolyte layers transports new electrons to the nanoparticle iron metal core also contributing to an additional negative charging of the iron metal core.

The hydroxyl ions that combine with the iron metal core will oxidize the iron and leave additional electrons on this metal core of the nanoparticle. These diffusion processes, involving hydroxyl ions, will also provoke the subsequent field and/or thermionic electron emission intra-particle breakdowns.

In humid air, an iron metal nanoparticle, being even uncharged with a surplus charge, can still be electrostatically hydrated by surrounding water vapor molecules because of its initial oxygen induced oxidative charging due to electron diffusion from the iron metal core to the growing semi-conductor FeO/Fe$_3$O$_4$ based shell (Fig. 11).

During electrochemical oxidation of the charged iron metal nanoparticle, quick-alternating electric dipole moments and huge potential differences can arise within the nanoparticle between the randomly arising local oxidative spots of positive charges on the outer surface of the nanoparticle and the corresponding uncompensated residual negative charges of the iron metal core.

Successive random "brownian" collisions of the nanoparticle with surrounding gas molecules will not only continuously and irregularly change the coordinates of the aerosol nanoparticle, but also, in a random manner, they will change the instantaneous values of the electric dipole moment of the nanoparticle because of the stochastic, "spotty", nature of the nanoparticle oxidation.

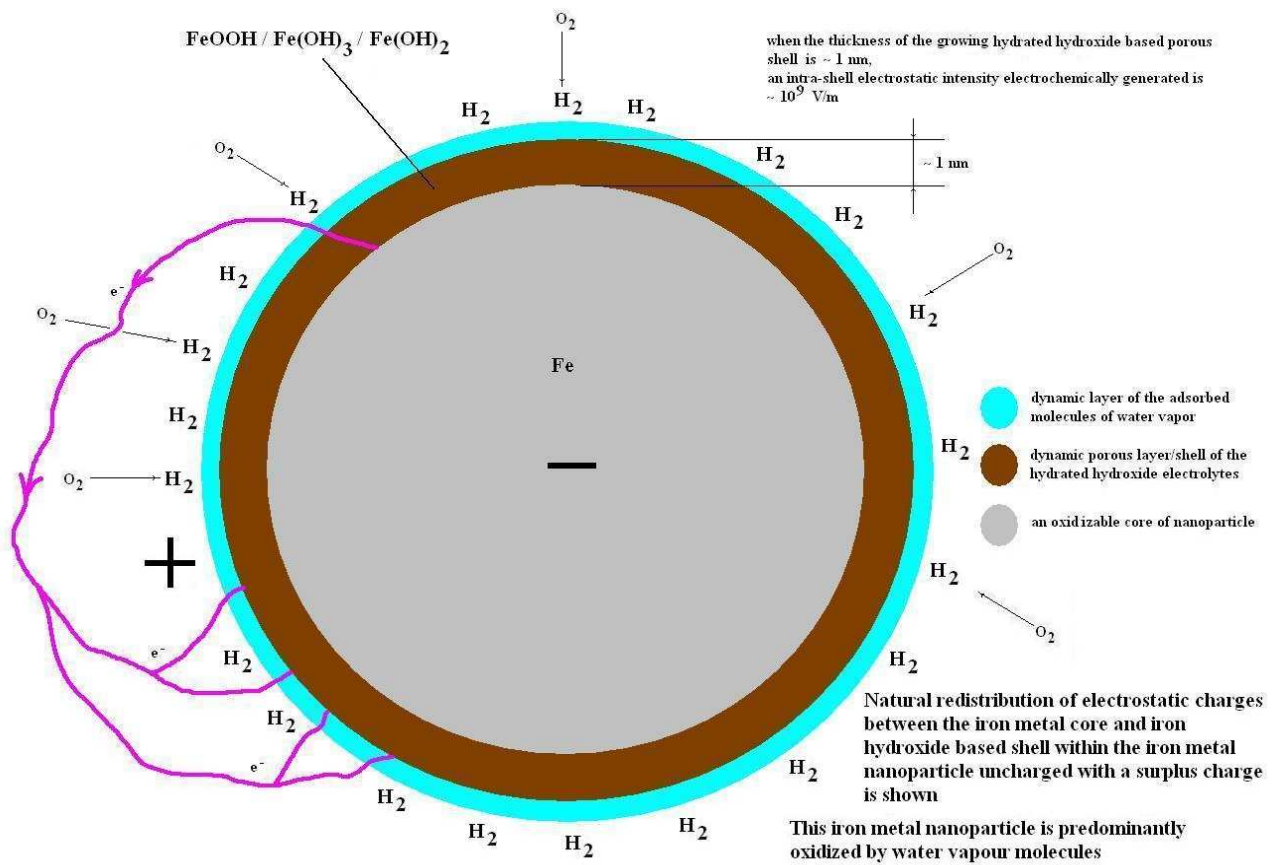

Figure 11. When electrostatically neutral metal aerosol particle is exposed to a gradual spotty oxidation in humid air, it is initially covered with a semi-conductor or dielectric metal oxide shell. Correspondingly, electron diffusion from the metal core into the semi-conductor or dielectric shell will charge the metal core with a positive charge, while the metal oxide shell of this particle will be spotty charged with a negative charge. Local spots of the negative charge will arise on the surface of this particle during its initial oxidation. These locally charged oxidized surface spots will attract surrounding polar molecules of water vapour from ambient air due to a charge-dipole interaction. So, the particle surface will be quickly and completely hydrated. Such electrostatic hydration of the nanoparticle surface can often transform the metal oxide semi-conductor or dielectric shell into the metal hydroxide or oxy-hydroxide electrolyte shell. In this case, at relatively low temperatures, preferential diffusion of the metal or hydroxyl ions within the hydrated electrolyte surface layers will form a new, opposite redistribution of electrostatic charges between the metal core (negatively charged now) and electrolyte shell (positively charged) with periodic relaxation of such electrochemically generated increasing electrostatic intensity by the field and thermoionic electron emission core-shell breakdowns.

The electrochemically generated fluctuating electric dipole moments, continuously arising within the nanoparticles-nanobatteries during oxidation, will cause a powerful electric dipole-dipole attraction between the separate aerosol nanoparticles-nanobatteries within a cloud of such nanobatteries. At the same time, the huge intra-particle core-shell voltage, electrochemically generated, will also cause repeating processes of the field and/or thermoionic electron emission breakdowns from the iron metal core to the outer surface of the nanoparticle during oxidation.

As these core-shell nanobatteries, spontaneously generated from the combustible iron metal aerosol nanoparticles during their oxidation, can be exposed to such continuously repeating relaxation processes (i.e. the intra-particle electron emission breakdowns), these periodically short-circuited aerosol nanoparticles-nanobatteries can be also described as pulsating aerosol current loops [20].

The strong alternating currents and corresponding magnetic moments generated within the periodically short-circuited aerosol nanobatteries can cause an additional powerful magnetic dipole-dipole attraction within a cloud of such nanobatteries.

Within a ball lightning cloud, despite the presence of unipolar charges on the surface of many aerosol nanoparticles-nanobatteries, both the electric and magnetic dipole-dipole attractions take place between the unipolar charged nanoparticles-nanobatteries. Both these alternating types of electromagnetic dipole-dipole attraction between charged aerosol nanobatteries contribute to significant dipole-dipole electromagnetic cohesion of the aerosol substance of such a cloud. (Certainly, the greater part of surplus electrostatic charges of the nanoparticles-nanobatteries within ball lightning will be probably localized on the peripheral aerosol nanoparticles due to mutual Coulomb repulsion between these unipolar charged nanoparticles and also due to a minimal air electroconductivity).

Reaction (44):

$$3\ Fe_{(nanoparticle)} + 6\ H_2O_{(vapour)} = Fe(OH)_2 + 2FeO(OH) + 4\ H_2\uparrow$$

is only slightly exothermal, with a moderate heat-evolution ~ 150 kJ per one mol of the produced $Fe(OH)_2$. However, a concomitant auto-ignition and combustion of the evolved hydrogen gas in ambient air could give considerable additional heat and water vapour:

$$4\ H_{2\ (gas)} + 2\ O_{2\ (gas)} = 4\ H_2O_{(vapour)} - 968\ kJ/mol \quad (50)$$

The repetitive processes of heating the nanoparticles-nanobatteries, caused by the repetitive processes of heat-evolution from reactions (36) and (40), can periodically and probably with a high frequency dehydrate the $FeOOH/Fe(OH)_2$ based surface electrolyte layers, which grow on these iron metal nanoparticles, and so this thermal nanoparticle dehydration will periodically inhibit the process of electrochemical oxidation of the charged iron metal nanoparticles with a temporary conversion of the iron hydroxide based surface electrolyte layers into the dehydrated, $Fe_3O_4$ based, semi-conducting surface layers of mixed iron oxides (Fig. 12).

However, a fast electrostatic re-hydration of the charged iron metal aerosol nanoparticle by surrounding polar molecules of water vapour in humid air will contribute to immediately cooling this combustible nanoparticle, and so the process of predominantly electrochemical, ion-mediated oxidation of the iron nanoparticle through its re-hydrated $FeOOH/Fe(OH)_2$ based electrolyte shell will be re-activated again. Probably, such self-oscillating processes of electrochemical oxidation of the charged combustible nanoparticles, continuously electrostatically re-hydrated by surrounding molecules of water vapour, could be quite high-frequency due to the extremely low heat capacity and the extremely high specific surface area of the aerosol nanoparticles.

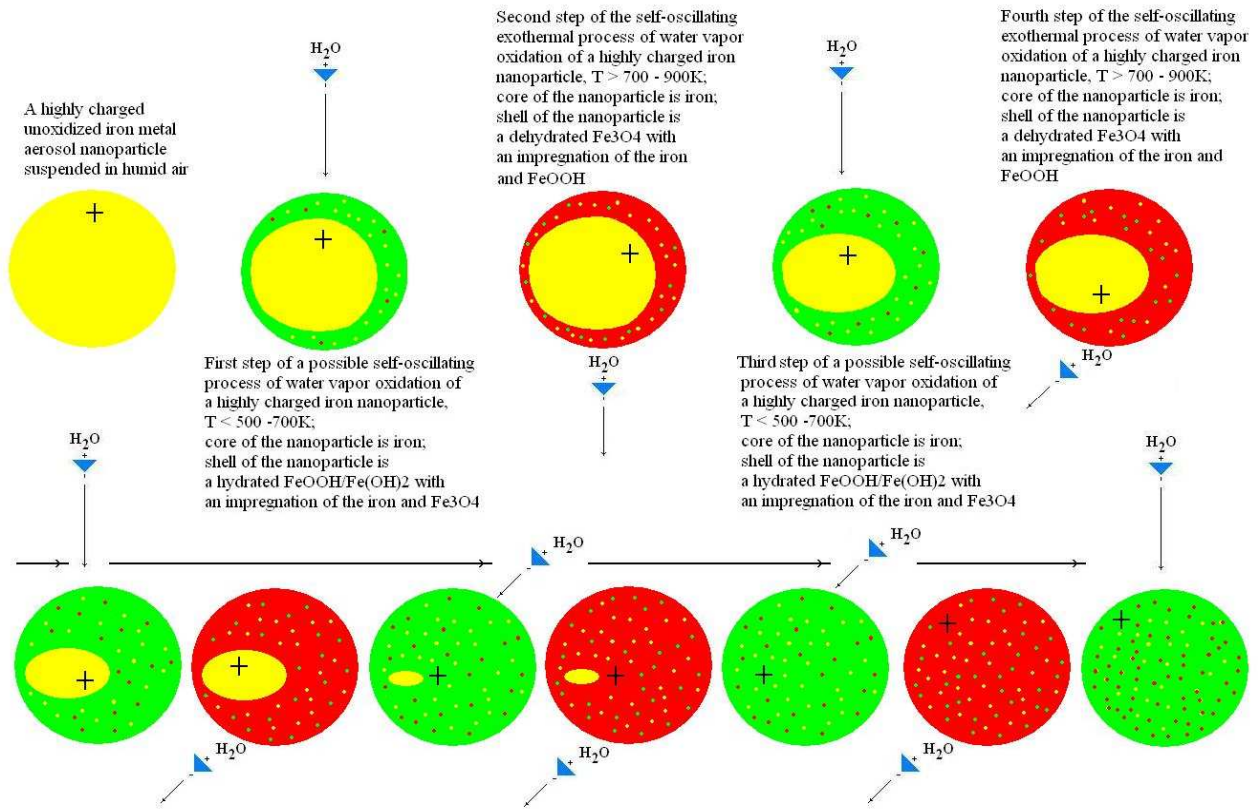

Figure 12. The stages of stepped exothermal oxidation of iron metal nanoparticles by surrounding polar molecules of water vapour successively alternate with stages of thermal dehydration of these nanoparticles to periodically convert the FeOOH / Fe(OH)$_2$ based electrolyte shells of the nanoparticles into the Fe$_3$O$_4$ / Fe$_2$O$_3$ based semi-conducting shells. Electrostatic oxidative adsorption of water vapour molecules forms a hydroxide based site on the charged nanoparticle surface. Such periodic oxidative electroadsorption of water vapour is accompanied by local evolving hydrogen gas. The hydrogen gas can auto-ignite in humid air, heating the nanoparticle and so again dehydrating it. These electrostatic oxidative hydration/ thermal dehydration alternating processes will repeat time and again. A final reaction product will consist of almost completely oxidized nanoparticles. The alternating stages of the nanoparticle's hydration/ dehydration are accompanied by a successive change of electrophysical characteristics of the core/shell junctions within the nanoparticles, from the iron metal / iron hydroxide electrolyte junctions to iron metal / iron oxide semi-conductor junctions, and then in the opposite direction. These periodic changes of the nanoparticle's surface characteristics can cause a successive intra-particle electron-ion transfer. Thus, the surface of the charged metal nanoparticle can continuously modify its composition and structure from metal oxide into metal oxy-hydroxide into metal hydroxide, then in the opposite direction etc., through such a self-oscillating thermocycling process of nanoparticle's humid air oxidation. An alternating electrostatic intensity of ~ one billion volts/metre will be generated inside the surface layers at each new stage of their repeating electrostatic hydration or thermal hydration. Correspondingly, the field or thermionic electron emission intra-particle breakdowns will occur at each stage of such re-increasing of core / shell electrostatic intensity.

These oxidative self-oscillating thermocycling processes can be partially (or totally) synchronized between all the charged aerosol nanoparticles-nanobatteries within ball lightning. Probably, such synchronization of the self-oscillating oxidative processes within ball lightning could also contribute to similar synchronization of the thermoionic electron emission processes of nanobatteries' short circuits with corresponding generation of powerful coherent radio frequency radiation by such a ball lightning cloud.

Within a ball lightning cloud, possibly, there is another mechanism in order that to synchronize an alternation of the stages of electrostatic oxidative hydration of nanoparticles-nanobatteries with the stages of electron emission breakdowns within these nanobatteries. The collective electron emission breakdowns of the aerosol nanobatteries within ball lightning are probably able to generate a wideband radio frequency radiation. At the same time, radio frequency radiation with resonant wavelengths, which are correlated with the specific diameter of the specific lightning ball, could become most strong. If a ball lightning nanobattery cloud is predominantly resonant self-oscillating radio frequency generator with the inner oxidative energy source, such model of the self-sustaining nanoparticle based resonant radio frequency aerosol generator can be considered as a similar and, at the same time, inverse model in relation to the theoretical and experimental mechanisms described in [5, 6, 37-41]. Within ball lightning, the collective short circuits within of trillions of the charged nanobatteries could generate the self-coordinated resonant oscillations (namely, the very high frequency or ultra high frequency or super high frequency oscillations according to the typical ball lightning diameters). Such resonant electromagnetic oscillations probably could in turn synchronize the successive intra-cloud processes of the periodic total dehydration, total heating and total thermoionic electron emission breakdowns within all or almost all the aerosol nanobatteries. A presence of such electromagnetic negative feedback within the cloud of the electrostatically charged and periodically short-circuited nanobatteries could help to understand a normal constancy of the ball lightning diameter during its life time; if certainly indeed ball lightning is a self-oscillating resonant radio frequency generator.

So, oxidation of the charged iron nanoparticles in humid air is a predominantly water vapour induced process. Correspondingly, this oxidative process can proceed predominantly through electrochemical, i.e. ion-mediated mechanism with formation of hydroxide or oxy-hydroxide or hydroxo-carbonate based nano-porous electrolyte layers on the charged nanoparticle's surface and evolving potentially inflammable hydrogen gas. Correspondingly, this humid air oxidative process can be accompanied by periodic auto-ignition of the evolved hydrogen gas and so this process can have a self-oscillating thermocycling character with significant temperature pulsations. With reference to the problem of ball lightning, it would also be interesting to find out whether minimally charged aluminium aerosol nanoparticles can be exposed to a similar strong electrostatic attack by accelerated high-reactive surrounding polar molecules of water vapour in humid air.

**9. Humid air oxidation of a spherical (~ 2 nm in diameter) aluminium metal aerosol nanoparticle charged with the minimum positive charge $Q = |e| = 1.6 \cdot 10^{-19}$ (C) at the aluminium melting-point of about 933 K.**

Let us again suppose that in humid air the mole fraction of molecules of water vapour, $n_W$, is ~ 0.02. Let us also suppose that at aluminium's melting-point of ~ 933 K (i.e. ~ 660°C), an average value of both activation energies, $E_{a\text{-}Al}$ (J/mol), for the competing processes of water vapour and oxygen gas induced oxidation of aluminium metal [i.e. an average value of the activation energy for the process of water vapour induced oxidation of an aluminium nanoparticle, $E_{aW\text{-}Al}$ (J/mol), and the activation energy for the process of oxygen gas induced oxidation of this nanoparticle $E_{aO\text{-}Al}$ (J/mol)] can be of ~ 200,000 (J/mol) [42]:

:

$$E_{a\text{-}Al} \approx E_{aW\text{-}Al} \approx E_{aO\text{-}Al} \sim 200,000 \text{ (J/mol)} \tag{51}$$

Thus, at $T \approx 933$ K:

$$kT \approx 1.38 \cdot 10^{-23} \cdot 933 = 1.29 \cdot 10^{-20} \text{ (J)} \tag{52}$$

$$\varepsilon_{a\text{-}Al} = E_{a\text{-}Al} / N_A \approx 200,000 / (6.02 \cdot 10^{23}) \approx 3.3 \cdot 10^{-19} \text{ (J)} \tag{53}$$

where $N_A \approx 6.02 \cdot 10^{23}$ – the Avogadro constant (mol$^{-1}$).

Consequently, in the case of humid air oxidation of the minimally charged aluminium aerosol nanoparticle at aluminium's melting-point:

$$\varepsilon_{a\text{-}Al} / kT = 25.6 \tag{54}$$

$$\varepsilon_{a\text{-}Al} / (kT + \varepsilon_1) = 3.3 \cdot 10^{-19} / (2.15 \cdot 10^{-20}) = 15{,}35 \tag{55}$$

Let us again recall that $\varepsilon_1 \approx 8.6 \cdot 10^{-21}$ (J) is the additional average kinetic energy acquired by the polar molecule of water vapour in the electrostatic field of a minimally charged nanoparticle at a distance of the mean free path from the nanoparticle.

Thus, the fraction of the reactive oxygen gas molecules, i.e. oxygen gas molecules possessing enough kinetic energy to climb the activation energy barrier of the oxygen gas induced oxidative reactions on the surface of the discussed aluminium nanoparticle at a temperature of about 933K, is:

$$\Delta N_O / N_O = \frac{2}{\sqrt{\pi}} \int_{25.6}^{\infty} \sqrt{\varepsilon'} e^{-\varepsilon'} d\varepsilon' \approx \frac{2}{\sqrt{\pi}} e^{-25..6} \left( \sqrt{25.6} + \frac{1}{2\sqrt{25.6}} \right) = 4.44 \cdot 10^{-11} \tag{56}$$

Consequently, the competing fraction of the reactive water vapour molecules possessing enough kinetic energy to climb the activation energy barrier of the water vapour induced oxidative reactions on the surface of the discussed minimally charged aluminium nanoparticle at a temperature of about 933K is:

$$\Delta N_H / N_H = \frac{2}{\sqrt{\pi}} \int_{15.35}^{\infty} \sqrt{\varepsilon'} e^{-\varepsilon'} d\varepsilon' \approx \frac{2}{\sqrt{\pi}} e^{-15..35} \left( \sqrt{15.35} + \frac{1}{2\sqrt{15.35}} \right) = 9.89 \cdot 10^{-7} \tag{57}$$

Thus, in this relatively high temperature case of oxidation of the minimally charged aluminium nanoparticle:

(a) the absolute quantity of the reactive non-polar molecules of oxygen gas, which is able to climb the activation energy barrier of the oxidative reactions on the surface of the discussed minimally charged nanoparticle at a temperature of about 933K, and which are contained in one mole of humid air:

$$\Delta N_{O-Al} = N_O \cdot 4.44 \cdot 10^{-11} = 1.26 \cdot 10^{23} \cdot 4.44 \cdot 10^{-11} = 5.6 \cdot 10^{12} \quad (58)$$

(b) the absolute quantity of the reactive polar molecules of water vapour, which is able to climb the activation energy barrier of the oxidative reactions on the surface of the discussed minimally charged nanoparticle at a temperature of about 933K, and which are contained in one mole of humid air:

$$\Delta N_{H-Al} = N_H \cdot 9.89 \cdot 10^{-7} = 1.2 \cdot 10^{22} \cdot 9.89 \cdot 10^{-7} = 1.19 \cdot 10^{16} \quad (59)$$

As one can see, the number of reactive water vapour molecules, which are able to oxidize the discussed minimally charged molten aluminium nanoparticle suspended in humid air at a temperature of ~ 933K is

$$\Delta N_{H-Al} / \Delta N_{O-Al} = 1.19 \cdot 10^{16} / 5.6 \cdot 10^{12} = 2.1 \cdot 10^{3} \quad (60)$$

times larger than the number of reactive oxygen gas molecules, which are able to oxidize the charged surface of this nanoparticle under such relatively high temperature conditions when the mole fraction of competing water vapour molecules, $n_W$, in humid air is ~ 0.02.

In this case, also an unexpectedly large (approximately two thousand times) difference is found between the number of reactive water vapour molecules, which are able to oxidize the minimally charged aluminium nanoparticle in humid air and the number of reactive molecules of oxygen gas, competitively participating in this humid air oxidation. This means that relatively few polar molecules of water vapour in humid air, being additionally electrostatically accelerated by their charge-dipole attraction towards the minimally charged aluminium nanoparticle, become practically the only essential air oxidant for the discussed minimally charged aluminium nanoparticle, and non-polar molecules of oxygen gas actually do not take part in humid air oxidation of this charged nanoparticle even at a temperature of aluminium melting of ~ 933K.

Thus, probably, that one of the main oxidative reactions on the surface of the minimally charged aluminium nanoparticle will be a water vapour induced exothermal reaction of oxidation of aluminium metal with formation of the hydrated, aluminium hydroxide Al(OH)₃ and/or aluminium oxy-hydroxide AlOOH based porous electrolyte surface layers with synchronous evolution of hydrogen gas:

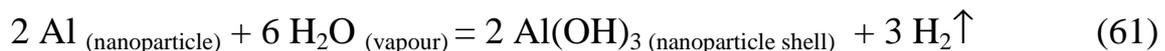
$$2\, Al_{(nanoparticle)} + 6\, H_2O_{(vapour)} = 2\, Al(OH)_{3\,(nanoparticle\ shell)} + 3\, H_2\uparrow \quad (61)$$

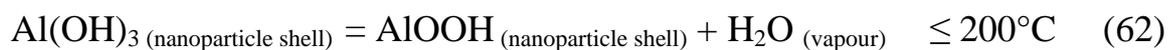
$$Al(OH)_{3\,(nanoparticle\ shell)} = AlOOH_{(nanoparticle\ shell)} + H_2O_{(vapour)} \quad \leq 200°C \quad (62)$$

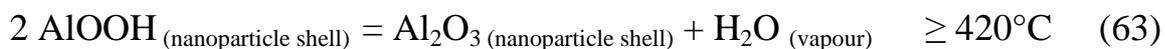
$$2\, AlOOH_{(nanoparticle\ shell)} = Al_2O_{3\,(nanoparticle\ shell)} + H_2O_{(vapour)} \quad \geq 420°C \quad (63)$$

In the case of the predominantly water vapour induced oxidation of the charged aluminium aerosol nanoparticles, the solid reaction products, growing on the surface of these nanoparticles mainly in the form of the thermolabile hydrated aluminium oxy-hydroxide based electrolyte layers saturated with hydrogen, will limit diffusion and direct oxidation of these aluminium nanoparticles by external neutral oxidizing species, for example such as oxygen gas, but at the same time, these electrolyte shells will effectively transport ionized species, such as $Al^{3+}$ or $Al(OH)^{2+}$ or hydroxyl ions, contributing to the preferential electrochemical mechanism of oxidation of the charged aluminium nanoparticles.

One of numerous possible oxidative reactions, which could take place directly on the hydrated aluminium hydroxide based surface of the charged aluminium metal nanoparticle-nanobattery, with the participation of the surrounding electrostatically accelerated water vapour molecules and the oxidizable aluminium-containing ions:

$$Al(OH)^{2+}_{(ion)} + 2e + 2H_2O_{(vapour)} \rightarrow Al(OH)_3 + H_2 \qquad (64)$$

Thus, in humid air, minimally electrostatically charged aluminium metal based aerosol nanoparticles might be oxidized by surrounding water vapour molecules rather than by non-polar oxygen gas molecules even at a temperature of ~ 933K. Such processes of intense electrostatic hydration and corresponding predominantly water vapor induced oxidation of the charged aluminium nanoparticles can automatically convert these nanoparticles into aluminium/air core-shell nanobatteries with the aluminium core anodes, the aluminium oxy-hydroxide based porous surface electrolyte nano-layers and outer surfaces of these electrolyte layers, which play the role of air-depolarized cathodes of these aluminium/air aerosol nanoparticles-nanobatteries.

As mentioned above, numerous electron emission breakdowns could probably occur within natural or artificial clouds consisting of the electrostatically charged metal, metalloid or carbon based nanoparticles-nanobatteries, for example within aluminium metal based ball lightning, during preferential electrochemical oxidation/ burning of such charged combustible nanoparticles in humid air. Even only partially synchronized intra-particle electron breakdowns within ball lightning consisting of trillions of aerosol nanoparticles-nanobatteries could generate a powerful radio frequency radiation, including microwave radiation.

It seems that a possible generation of more or less intense radio frequency radiation over the range of ~ 0.1 gigahertz up to ~ 20 gigahertz from burning aerosol clouds, which consist of artificially or naturally charged combustible particles, could be relatively easily found in corresponding aerosol combustion experiments. Probably, a condensed disperse phase of such radio frequency / microwave emitting slowly burning aerosol clouds could consist of bipolar or unipolar charged nano or submicron combustible particles, e.g. such as aluminium-, or iron-, or magnesium-, or zirconium based nanoparticles, as well as soot- or silicon based nanoparticles with an electrolyte-modified surface.

In standard humid air, predominantly electrochemical burning of such electrostatically charged nanoparticles could be accompanied with intense thermoemission and photoemission surface processes, automatically sustaining spontaneous electrostatic charging of these burning nanoparticles.

In particular, the formation of electrochemically burning aerosol clouds, capable of generating powerful radio frequency radiation, can be spontaneously realized in various hypervelocity flight processes, e.g. in a troposphere movement of iron-nickel based meteors or in a flight of hypervelocity rockets. For example, during the troposphere movement of hypervelocity rockets, aerosol clouds consisting of the combustible short-circuited nanoparticles-nanobatteries can be produced either by high-temperature ablation of heat-insulating materials or by incomplete combustion of carbon based fuels. The presence of combustion-generated water vapour, highly charged combustible soot nanoparticles and additional electrolyte impurities in plasma-condensation trails of hypervelocity rockets can probably contribute to activation of a charge-catalyzed water vapour induced electrochemical oxidation of the highly charged soot based nanoparticles-nanobatteries with a resultant generation of powerful radio-interference, including microwave noise, which could, for example, interrupt radio telemetry.

## 10. Some final remarks

Clearly, the discussed effect of the charge-catalyzed water vapour induced oxidation of unipolar and/or bipolar, electrostatically or electrodynamically, charged aerosol nanoparticles/ substrate-integrated nanostructures can take place during the processes of humid air oxidation of various combustible nanoobjects, not only metal or metalloid ones, as the large additional kinetic energy, acquired by surrounding polar molecules of water vapour in the electrostatic field of the charged nanoparticles at a distance of their mean free path from the nanoparticles, is independent of the materials constituting these nanoobjects. A change in polarity of the surplus charge of the combustible nanoparticles is also not able to change the value of the additional kinetic energy acquired by surrounding polar gas molecules in the electrostatic field in immediate proximity to the charged nanoparticles.

Correspondingly, combustible particles charged with equal, either negative or positive, electrostatic charges can be equally intensively attacked by surrounding water vapour molecules accelerated in the electrostatic field of these nanoparticles, irrespective of the polarity of the nanoparticle's surplus charges.

In many cases, in humid air, such an intense charge-catalyzed water vapour induced oxidation of the charged metal or metalloid nanoparticles can mainly proceed through the electrochemical mechanism due to the preferential growth of the hydrated hydroxide based dynamic electrolyte surface layers, re-saturated with hydrogen gas. These thermally unstable, continuously renewed and again decomposed, passivating surface electrolyte layers can significantly inhibit further oxidation of combustible nanoparticles by gas phase neutral oxidizing species. At the same time, these surface electrolyte layers will stimulate the relatively low-temperature process of the ion-mediated oxidation of the combustible nanoparticles. A time-average temperature of such electrochemical oxidation of the combustible nanoparticles can be relatively low, probably ranging from ~ 300 to 900°C, particularly taking into account the above described scenario of the self-oscillating electrochemical oxidation with the alternating thermocycling processes of oxidative electrostatic hydration and thermal flame dehydration of the combustible particles.

According to [20], when an average diameter of the soot based aerosol nanoparticles-nanobatteries constituting a soot based ball lightning cloud is ~ 100nm and when a total mass of this condensed phase of the ball lightning cloud is ~ 4 grammes, then the average number of such carbon/air nanoparticles-nanobatteries can be of ~ $10^{15}$ in this ~ 20 cm diameter ball lightning cloud. If a net electrostatic charge of such 'average' ball lightning ranges from ~ 0.1 to 1 microcoulomb, i.e. from ~ $6.2 \times 10^{11}$ to ~ $6.2 \times 10^{12}$ elementary charges, then large aerosol nanoaggregates consisting of ~ 200-2.000 such nanoparticles can be charged only with a single surplus mobile ion per one nanoaggregate. At the same time, larger, e.g. micrometre-sized carbon based particles-batteries or denser metal (e.g. iron) based submicron-sized particles-batteries could be charged with a single surplus mobile ion per one large compact submicron iron metal particle.

In any case, it is evident that combustible aerosol particles constituting ball lightning can only have ~ one surplus ion per one aerosol particle on average. Correspondingly, within ball lightning, the charge-catalyzed water vapor induced oxidation of combustible aerosol particles will practically always proceed through the step-by-step successive mechanisms, which are illustrated in the Figures 2-5. Equally, it is clear that a high hydrophility of the oxidized surface of these combustible nanoparticles (the high hydrophility is a typical property of the overwhelming majority of metal oxides, metal oxy-hydroxides and metal hydroxides) could enable a fast redistribution of locally electroadsorbed water vapour molecules over all the reacting surface of such large, submicron or micrometre-sized, either aggregated or compact, combustible aerosol particles.

As one can see, a lot of the combustible aerosol particles constituting the ball lightning cloud are, virtually, uncharged with a surplus charge. At the same time, within ball lightning these particles are strongly electrostatically polarized, i.e. they are charged with local bipolar charges, and so these oxidizable particles can still be exposed to an intense charge-catalyzed water vapour induced oxidation.

For the sake of brevity, a lot of important additional aspects can not be considered in this paper. Although certain additional aspects, such as a temperature-dependent intra-particle core-shell jumping of the nanoparticle's mobile surplus charges, an influence of the polarity, mobility and intra-particle location of the surplus ions on the rate of electrochemical oxidation of the combustible nanoparticles, a surplus charge-dependent intense 'anodic' oxidation of multiply negatively charged combustible nanoparticles or, on the contrary, a surplus charge-dependent inhibition of oxidation of multiply positively charged combustible nanoparticles ('cathodic protection') – the hypothetic effects, which are somewhat similar to those described in [43, 44], could be important to better understand ball lightning phenomenon, however all these aspects fall outside the limits of this necessarily limited paper.

Various possible variants of electrochemical redistribution of charges inside metal aerosol nanoparticles - short-circuited metal/air aerosol nanobatteries, minimally charged and hence actively electrostatically hydrated in humid air, are shown in the Figures 13-18.

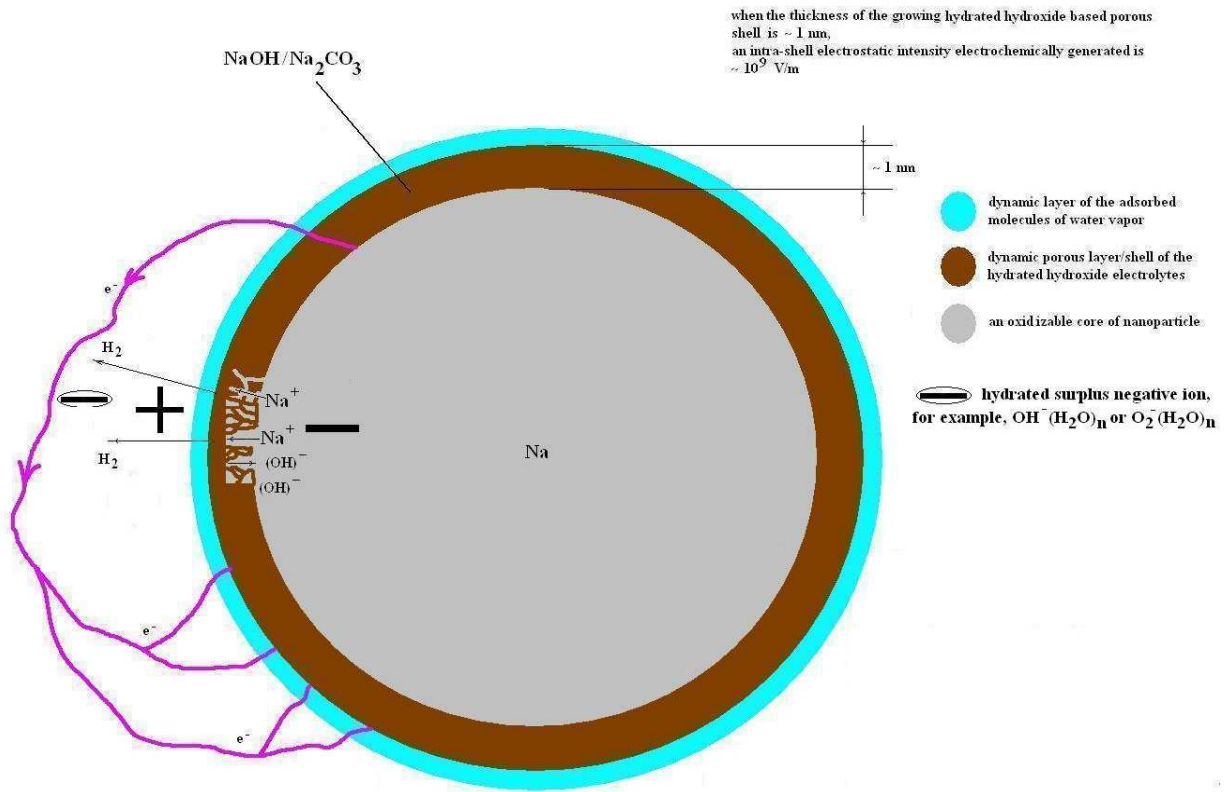

Figure 13. In humid air, negatively charged sodium metal based nanoparticles or their aggregates with soot carbon nanoparticles are continuously re-electrohydrated by surrounding molecules of water vapour, and so they can be spontaneously transformed into the sodium/air or composite Na/carbon/air core-shell nanobatteries, which are periodically short-circuited by the intra-particle core-shell electron emission breakdowns. Probably, the small ball-shaped clouds of such sodium/air core-shell nanoparticles-nanobatteries could be generated by high-voltage pulse-arc electrolysis from the $NaHCO_3$ containing water solutions in the experiments [4, 7]. At an initial stage, the sodium metal and hydrogen gas could be synchronously produced on the hot carbon cathode. In these experiments, milligrammes of the fresh-reduced electrolysis-generated sodium metal could be immediately evaporated by the pulse arc discharge. Then, the evaporated sodium metal can be condensed in the form of the small cloud of the negatively charged sodium metal nanoparticles in the local hydrogen gas based reducing atmosphere. In humid air, predominantly electrochemical oxidation of the highly charged and continuously electrohydrated sodium nanoparticles-nanobatteries could probably take place in these specific experiments with the $NaHCO_3$ containing water solutions and carbon cathode, while similar ball lightning clouds consisting of the negatively charged sodium-, or calcium-, or iron-, or aluminium- metal based nanoparticles, which are spontaneously converted into metal/air or carbon/air core-shell nanobatteries, could be generated by both the pulse electrolysis and arc evaporation of carbon or metal cathodes in the earlier similar experiments [1, 2].

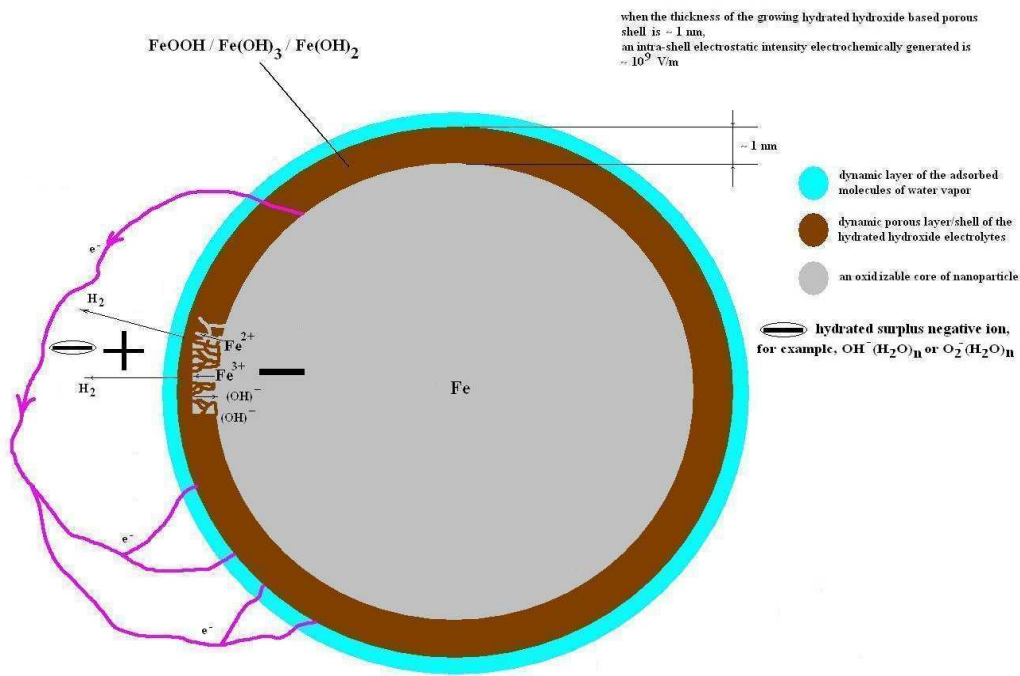

Figure 14. In humid air, negatively charged iron metal nanoparticles are continuously re-electrohydrated by surrounding molecules of water vapour, and so they can be spontaneously transformed into the iron/air core-shell nanobatteries, which are periodically short-circuited by the intra-particle electron emission breakdowns.

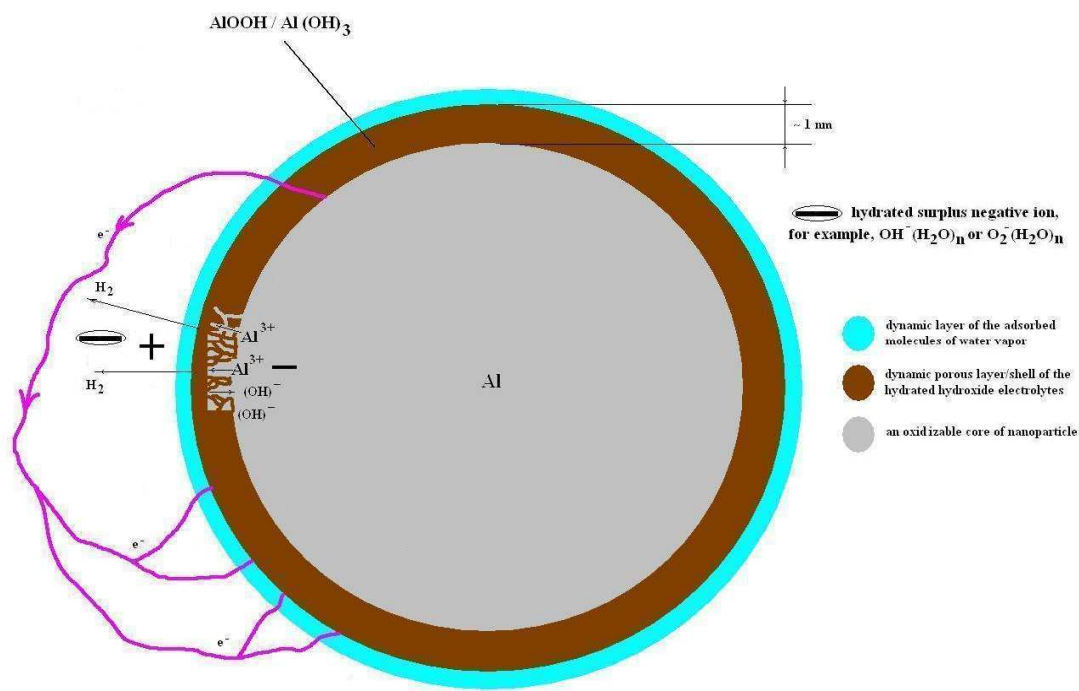

Figure 15. Similarly, in humid air, negatively charged aluminium nanoparticles can repeatedly be electrohydrated by surrounding molecules of water vapour, and so they can be spontaneously transformed into the aluminium/air core-shell nanobatteries, which are periodically short-circuited by the intra-particle electron emission breakdowns.

The simplest two-particle aerosol nanoaggregate can, for example, consist of:
(a) an oxidable metal nanoparticle, e.g. such as an iron, or aluminium, or tungsten, or molybdenum, or silicon nanoparticle, (b) a carbon/soot nanoparticle, and
(c) a hydrated oxide-hydroxide/hydroxide boundary layer (green). In humid air, the uncharged oxidizable metal or metalloid nanoparticle (yellow) is only slightly
covered by a hydrated oxide-hydroxide/hydroxide based shell (green), consisting of such components as, e.g.:
$FeOOH$, or $AlOOH$, or $H_2WO_4$, or $H_2MoO_4$, or $H_2SiO_3$.
If this metal/carbon nanoaggregate is uncharged with an additional charge, i.e. if $Q = 0 (C)$, a surface hydration of the nanoaggregate by water vapor molecules
from ambient air is minimum, and the process of an electrochemical charge separation within such a neutral nanoaggregate-nanobattery is weakly marked.

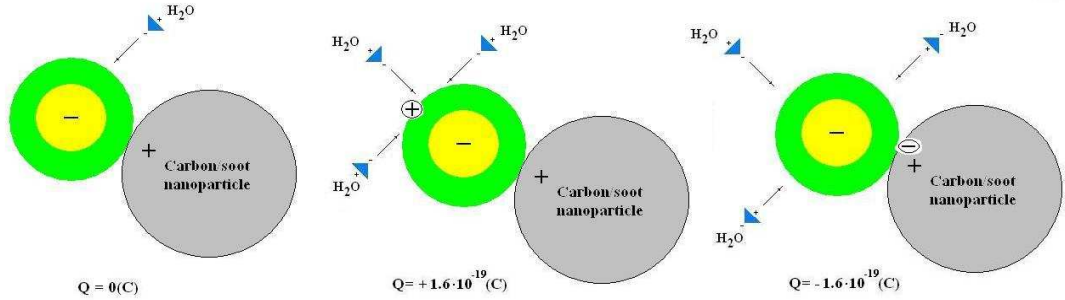

In an alternative case, if a metal/carbon nanoaggregate is charged, an electrostatic hydration of such a charged nanoaggregate by water vapor molecules
from ambient air may be substantial due to a charge-dipole attraction between the charged nanoaggregate and surrounding polar molecules of water vapor.
Thus, processes of the electrochemical oxidation of the metal nanocomponent and internal electrochemical redistribution of charges within the charged
aerosol nanoaggregate-nanobattery can be significantly activated by its electrostatic hydration.
This situation recalls the filling of a dry accumulator - charged aerosol metal/carbon/air nanoaggregate-nanobattery - with a liquid electrolyte - superheated
electrostatically accelerated molecules of water vapour that are continuously or periodically absorbed on the surface of the charged nanoaggregate-nanobattery.
The charged metal/carbon/air nanoaggregates-nanobatteries are evolving hydrogen gas and are periodically short-circuited by an electron emission from a
metal anode to a carbon cathode of these aerosol nanobatteries.

Figure 16. Charged aggregated aerosol nanocomposites, for example aggregated aerosol nanobatteries or aggregated aerosol nanothermites, are continuously electrostatically re-hydrated in humid air. This can result in a periodic transformation of the metal oxide based dielectric or semi-conductor interphase barriers into the metal hydroxide based electrolyte interphase barriers, with the corresponding periodic re-activation of the electrochemical and electron emission discharge processes in the aggregated aerosol nanocomposites.

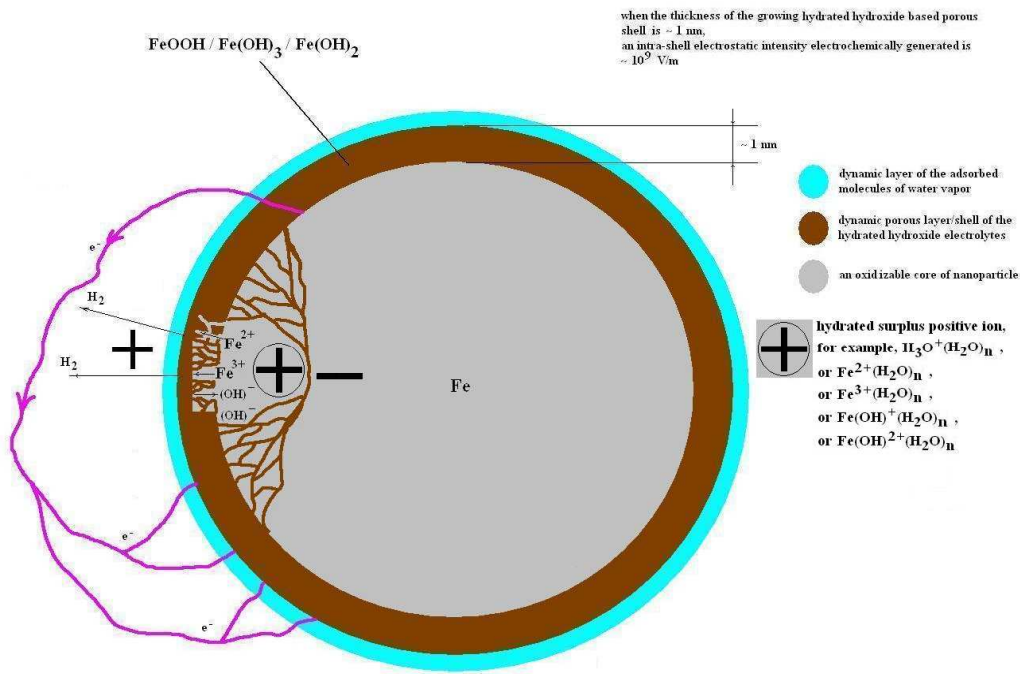

Figure 17. Positively charged iron nanoparticles can equally be continuously re-electrohydrated by surrounding molecules of water vapour, and so they can be transformed into the iron/air core-shell nanobatteries, periodically short-circuited by the intra-particle electron emission breakdowns.

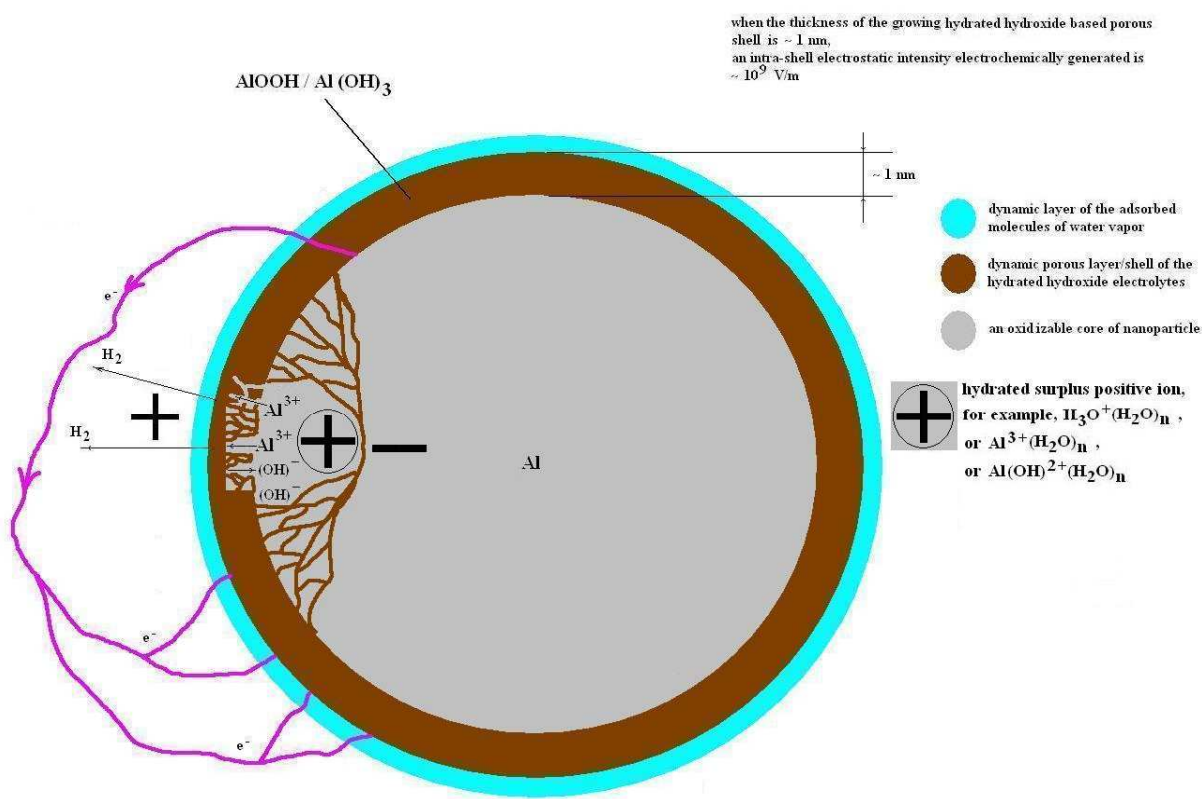

Figure 18. Positively charged aluminium nanoparticles can equally be electrohydrated by surrounding molecules of water vapor, and so they can be transformed into the aluminium/air core-shell nanobatteries, periodically short-circuited by the intra-particle electron emission breakdowns.

The additional kinetic energy of the electrostatic acceleration acquired by polar gas molecules near to a charged nanoparticle is independent not only of the specific material components of the charged nanoparticle and the polarity of the nanoparticle's surplus charge, but also of the presence and type of chemical reactions on the surface of the nanoparticle.

This means that the substantial additional kinetic energy of $\geq 10^{-20}$ (J), which can be acquired by various electrostatically accelerated polar gas molecules in immediate proximity to charged aerosol nanoparticles (or charged surface-integrated nanostructures), could contribute to various heterogeneous reactions, not only oxidative ones, on the charged surface of such particles or surface-integrated structures which are in contact with any potential polar gas phase reactants.

In such cases, electrostatic charges (solvated ions or electrons), either adsorbed by these nanoparticles/ nanostructures or induced on their surface by external fields, could probably act as local catalysts of various heterogeneous reactions.

Of course, the successive process of the charge-catalyzed predominantly water vapour induced electrochemical oxidation can take place in a humid atmosphere not only in relation to highly charged nano or micrometre sized combustible aerosol particles or substrate-integrated structures, but also in relation to certain highly charged macroscopic objects, for example, such as the corona-forming highly porous rusty iron/carbon based electrodes, where a secondary electrospraying effect, which we earlier named 'electrostatic metal dusting corrosion', could be caused by a local charge-catalyzed water vapour induced oxidation of both carbon and iron metal components of corona-forming corrodable electrodes, with an additional local generation of water gas. In this case, the charge-dipole attraction between the charged porous surface of the iron/carbon based high-voltage electrodes and the surrounding molecules of water vapour could again play the role of an electrostatic water pump to selectively transport and accelerate the water vapour polar molecules, but not the oxygen gas molecules, towards the charged oxidizable surface of the high-voltage electrodes.

Normally, uncharged pure carbon based nanomaterials, for example soot nanoparticles are quite inert in humid air. However, it seems that highly charged carbon based nanoparticles can be much more reactive. One of the possible direct reactions of the surrounding electrostatically accelerated molecules of water vapour with the charged carbon based aerosol nanoparticle can be the so-called water gas endothermic reaction:

$$C_{(nanoparticle)} + H_2O_{(vapour)} = CO + H_2 + 131.3 \text{ kJ/mol} \tag{65}$$

Though this reaction is able to generate a combination of the two high-calorific fuel gases, this first stage of process of the water vapor induced oxidation of the carbon based nanoparticle absorbs a lot of heat.

Alternative reactions of the charged carbon nanoparticle with the non-polar and, accordingly, electrostatically non-accelerated molecules of ambient oxygen gas are exothermal reactions:

$$2 C_{(nanoparticle)} + O_2 = 2 CO - 221 \text{ kJ/mol} \tag{66}$$

with carbon monoxide as the predominant product at a temperature above 800 °C, and

$$C_{(nanoparticle)} + O_2 = CO_2 - 393.5 \text{ kJ/mol} \tag{67}$$

with carbon dioxide as the predominant product at a temperature ~ 600-700 °C.

Intense electrostatic hydration of the oxygen gas pre-oxidized carbon based red hot nanoparticles by the surrounding polar molecules of water vapour will contribute both to a temporary cooling of these nanoparticles and to a pulse evolution of the secondary inflammable gases, i.e. hydrogen gas and carbon monoxide gas. In ambient humid air, the heat released from subsequent auto-ignition of these high-calorific fuel gases:

$$2 H_{2 \text{ (gas)}} + O_{2 \text{ (gas)}} = 2 H_2O_{(vapour)} - 484 \text{ kJ/mol} \tag{68}$$

$$CO + H_2O_{(vapour)} = CO_2 + H_2 - 41.2 \text{ kJ/mol} \tag{69}$$

$$2 CO + O_2 = 2 CO_2 - 566 \text{ kJ/mol} \tag{70}$$

is again able to dehydrate and make the electrostatically charged carbon based nanoparticles red-hot. This dehydration is able to re-start the relatively high-temperature process of oxygen gas induced oxidation of these electrostatically charged carbon based nanoparticles. The high heat emission of these nanoparticles and their relatively low reactivity in normal humid air will again contribute to the fast cooling of these nanoparticles and to the re-starting of their electrostatic hydration and their relatively low-temperature water vapour induced oxidation. Thus, such a prolonged self-oscillating thermocycling process of the fast alternation of oxygen gas and water vapour induced oxidation of the highly charged soot nanoparticles can probably take place in humid air.

Probably, pure water based dynamic shells, consisting only of the electroadsorbed epithermal water vapour molecules, electrostatically attacking the surface of the charged soot nanoparticles, are not able to convert these charged carbon nanoparticles into carbon/air nanobatteries. The pure superheated water is still a weak surface electrolyte.

It is possible that additional strong electrolyte surface impurities, such as sodium hydroxide/ sodium carbonate or potassium hydroxide/ potassium carbonate or calcium hydroxide/ calcium carbonate or barium hydroxide/ barium carbonate aerosol impurities, co-condensed and aggregated along with soot nanoparticles, could contribute to a conversion of such molten salt electrolyte coated electrostatically charged soot nanoparticles into the carbon/air aerosol nanobatteries in humid air.

The epithermal effects of the intense electrostatic acceleration and corresponding substantial activation of various polar gas phase molecules, for example such as gas phase molecules of hydrogen peroxide, nitric acid, nitrogen monoxide, carbon monoxide, ozone, sulphuric acid, ammonia, hydrogen sulphide, hydrochloric acid, as well as many organic polar molecules, which take place in immediate proximity to naturally or artificially electrostatically charged oxidizable nano-objects, can perhaps be successfully used in many highly varied nano-technological applications, for example in those such as heterogeneous catalysis, heterogeneous polymerization of polar monomers (e.g. methyl methacrylate, styrene etc) on the surface of highly charged nanoscale catalysts, in exhaust systems with the purpose of a secondary combustion of soot nanoparticles artificially highly pre-charged, either with the help of a high-frequency corona discharge, or by bipolar direct current corona discharge, either by contact, or microwave or plasma charging. Such electrostatic or electrodynamic catalysis of the oxidation/ combustion processes with an intense high-frequency bipolar charging of solid or liquid fuel particles in engines or power generating plants could also be of practical interest.

Generally speaking, similar intense charge-dipole attacks of electrostatically extra accelerated polar gas molecules, directed against adjacent highly charged nano-objects, can probably be a widespread phenomenon in nature. In particular, such electrostatic attacks of the polar gas molecules (e.g. nano-bubble contained highly polar gas molecules of hydrogen peroxide, hydrogen sulphide, hydrochloric acid, nitrogen monoxide, carbon monoxide etc), directed against naturally or artificially, electrostatically or electrodynamically, charged nano-bio-objects, for example, such as lipoproteins open to injury by oxidation or DNA's open to injury by mutations that can be generated due to possible high-energy collisions DNA's based biopolymers with the surrounding charge-dipole extra accelerated epithermal polar gas molecules, could also play the important role in both natural biological processes and potential bio-medical uses.